\documentclass{article}
\usepackage{cachetune_preprint,times}
\usepackage[T1]{fontenc}
\usepackage[utf8]{inputenc}
\usepackage{amsmath,amssymb}
\usepackage{graphicx,adjustbox}
\usepackage{booktabs,tabularx,array}
\usepackage{algorithm}
\usepackage{url}
\usepackage{hyperref}
\usepackage{microtype}
\usepackage{placeins}
\usepackage{capt-of}
\usepackage{needspace}
\hypersetup{hidelinks, pdftitle={Adaptive KV Cache Reuse for Fast Long-Context LLM Serving},pdfauthor={Fei Li, Song Liu, Yan Liu, Jinhua Cui, Shiqiang Nie, Jinyu Wang, Weiguo Wu}}
\title{Adaptive KV Cache Reuse for\\Fast Long-Context LLM Serving}
\author{%
  {\normalsize\bfseries Fei Li\textsuperscript{1}, Song Liu\textsuperscript{1},
  Yan Liu\textsuperscript{1}, Jinhua Cui\textsuperscript{2},\\
  Shiqiang Nie\textsuperscript{1}, Jinyu Wang\textsuperscript{1},
  Weiguo Wu\textsuperscript{1}}\\[5pt]
  {\small\textsuperscript{1}School of Computer Science and Technology,\\
  Xi'an Jiaotong University, Xi'an, China}\\[3pt]
  {\small\textsuperscript{2}School of Computer Science and Technology,\\
  Huazhong University of Science and Technology, Wuhan, China}\\[5pt]
  {\footnotesize\texttt{lifei@stu.xjtu.edu.cn}, \texttt{liusong@mail.xjtu.edu.cn}}\\
  {\footnotesize\texttt{gmliuyan@stu.xjtu.edu.cn}, \texttt{csjhcui@gmail.com}}\\
  {\footnotesize\texttt{shiqiang.nie@xjtu.edu.cn}, \texttt{jinyu.wang@xjtu.edu.cn}}\\
  {\footnotesize\texttt{wgwu@xjtu.edu.cn}}%
}


\begin{document}
\raggedbottom
\maketitle
\begin{abstract}
Key-Value (KV) Cache reuse reduces prefill computation for long-context large language models (LLMs), but reusing independently encoded text chunks in non-prefix contexts degrades generation quality because cross-chunk attention is missing. Selective recomputation can mitigate this problem, but it requires both selecting effective tokens for recomputation and preventing the costs of recomputation and cache loading from offsetting the benefits of reuse. We observe that the frequency-domain structure of KV provides an effective signal for identifying tokens that help restore cross-chunk dependencies. Based on this observation, we propose CacheTune, a frequency-guided method for non-prefix KV Cache reuse. CacheTune analyzes independently encoded KV Chunks offline to generate recomputation indices shared across layers, recomputes the selected KVs in the new context, and reuses the remaining cache, without additional training or online importance analysis. The same indices also determine which KVs must be loaded, so the volume of cache transfers decreases as the recomputation ratio increases, while allowing transfer to overlap with computation. CacheTune further calibrates the recomputation ratio based on hardware computation costs and cache access costs to accommodate different storage tiers. Across mainstream models and long-context tasks, CacheTune achieves a $3.72$--$4.86\times$ speedup in time-to-first-token (TTFT) and $3.93$--$6.21\times$ the throughput of full recomputation while maintaining comparable generation quality. With SSD/HDD caches, the TTFT speedup still reaches $2.34$--$2.36\times$.
\end{abstract}

% Source Word paragraph 5

\section{Introduction}\label{sec:1}
% Source Word paragraph 6
Large language model (LLM) serving is gradually shifting from short conversations to long-context applications such as retrieval-augmented generation (RAG) \citep{lewis2020retrieval}, long-document question answering \citep{bai2024longbench}, and agent reasoning \citep{DBLP:conf/iclr/YaoZYDSN023}. In these scenarios, input prompts are typically assembled dynamically from retrieved documents, conversation histories, or reusable knowledge segments. Prefill must process the entire context, and its computational cost grows rapidly with input length. The resulting time-to-first-token (TTFT) has become a major performance bottleneck in long-context serving \citep{gim2024prompt,yao2025cacheblend}. The same document or knowledge segment may appear in multiple requests. Encoding them anew each time also repeats the corresponding Key-Value (KV) generation, further increasing the prefill workload.
%%multi-document summarization \citep{fabbri2019multi}

% Source Word paragraph 7
A natural way to eliminate redundant prefill computation is to materialize the KV Cache generated from reusable text segments as independently cacheable units, hereafter called KV Chunks, and reuse them across requests \citep{ye2024chunkattention,liu2023cachegen}. Prefix caching follows this approach and has been adopted in systems such as vLLM \citep{kwon2023efficient} and SGLang \citep{zheng2024sglang}. However, it requires the KV Chunks to form a strict prefix of the new request. In RAG and multi-document workloads, reusable content typically consists of independent text chunks whose number, order, and positions change dynamically, preventing prefix caching from fully exploiting these repeated data.

% Source Word paragraph 8
Allowing independent KV Chunks to be reused at arbitrary positions can improve cache hit rates, but substantially degrades generation quality. The fundamental reason is that a Transformer's causal self-attention mechanism makes a token's KV depend not only on its own content but also on all preceding context. Text chunks encoded independently in advance lack the cross-chunk attention that should have been established in the complete prompt. Their KV therefore represents a context-mismatched materialized state relative to the current global context. We refer to this inter-chunk attention as \textbf{cross-attention}. Directly concatenating these chunk caches causes the resulting discrepancies to propagate into subsequent attention computations and the final generated output. Non-prefix KV Cache reuse must therefore repair the invalidated contextual dependencies \citep{DBLP:conf/icml/HuHWWHZFCS025}.%%,liu2026cacheslide

% Source Word paragraph 9
Selecting important KV for recomputation is a key approach to restoring cross-attention. Existing recomputation methods organize cross-attention repair based on online KV deviations, chunk boundary positions, or learned link tokens \citep{yao2025cacheblend,DBLP:conf/icml/HuHWWHZFCS025,DBLP:journals/corr/abs-2502-16002}. These methods respectively rely on online analysis, fixed positional priors, or additional training. We consider another question: can effective selection signals be extracted solely from an independent chunk's KV, so that the KV indices to be recomputed can be determined when the cache is generated and reused across requests? We observe that KV energy along the token dimension is concentrated in low-frequency components. After the low-frequency signal is reconstructed at the original positions, tokens selected by their magnitude contribute more to restoring attention distributions and task quality under the same recomputation budget. This observation provides a basis for an offline token importance ranking independent of the current query. In addition, recomputation index selection must account for cache access costs. As reusable KV Caches exceed GPU memory capacity and expand to lower storage tiers or even external caching systems \citep{lin2026kvdrive,gao2024cost}, cache transfer and online recomputation become two competing latency paths. The computational savings from KV reuse may be offset by additional cache-loading overhead \citep{qin2024mooncake,liu2025lmcache}. Efficient reuse therefore requires dynamically balancing cache transfer and online recomputation.
%%10.14778/3796195.3796212, ,lee2024infinigen

% Source Word paragraph 10
Based on these observations, we propose CacheTune, which manages the repair, transfer, and reuse of non-prefix KV Caches across storage tiers while preserving generation quality. During cache generation, CacheTune performs a one-time offline frequency-domain analysis of each independently encoded KV Chunk to identify the tokens most critical to global semantic recovery, enabling targeted repair of cross-chunk attention. To hide cross-tier I/O overhead, CacheTune uses an asynchronous pipeline with multiple CUDA streams to overlap sparse KV transfer, online recomputation, and deferred positional-encoding recovery. Building on this pipeline, a hardware-aware search for the recomputation ratio adaptively balances computation and data movement across heterogeneous storage tiers. CacheTune thus enables high-quality non-prefix KV reuse with accuracy close to full recomputation and substantially reduces TTFT across different storage configurations. Our contributions are summarized as follows:

% Source Word paragraph 11
\begin{itemize}
\item We propose a frequency-guided method for KV importance modeling and selective recomputation in non-prefix KV Cache reuse. Using the intrinsic frequency-domain structure of independently encoded KV Chunks, the method identifies offline the KV indices most critical to restoring cross-chunk semantic dependencies, effectively recovering the inter-chunk cross-attention missing in non-prefix reuse at a low recomputation cost.

% Source Word paragraph 12
\item We propose an algorithm--system co-design framework that translates frequency-guided indices into a complementary recomputation and reuse plan. CacheTune transfers only the KV that does not require recomputation, asynchronously overlaps sparse KV transfer, selective recomputation, and deferred positional-encoding recovery, and further adjusts the recomputation ratio to the computation and I/O characteristics of the target cache tier. By coordinating data movement across cache tiers with recomputation on the compute device, CacheTune reduces TTFT in multi-tier caching environments.

% Source Word paragraph 13
\item We evaluate CacheTune across multiple models, long-context tasks, and KV Cache storage tiers. While maintaining generation quality close to full recomputation, CacheTune achieves a $3.72$--$4.86\times$ TTFT speedup and $3.93$--$6.21\times$ the serving throughput, offering a better quality--latency trade-off than state-of-the-art baselines. With I/O-bound SSD and HDD cache pools, CacheTune still achieves a $2.34$--$2.36\times$ TTFT speedup.

\end{itemize}

% Source Word paragraph 14

\section{Related Work}\label{sec:2}
% Source Word paragraph 15
Prefix caching systems, including vLLM, SGLang, and ChunkAttention, reuse only matching prefixes \citep{kwon2023efficient,zheng2024sglang,ye2024chunkattention}. Non-prefix methods relax this constraint through modular caching or cross-chunk repair. PromptCache uses predefined prompt schemas \citep{gim2024prompt}; CacheBlend selects tokens with large KV deviations through online first-layer recomputation \citep{yao2025cacheblend}; EPIC repairs fixed attention-sink positions \citep{DBLP:conf/icml/HuHWWHZFCS025}; and KVLink introduces learned link tokens \citep{DBLP:journals/corr/abs-2502-16002}. CacheSlide addresses positional drift when reusable segments preserve their relative order \citep{liu2026cacheslide}. CacheClip \citep{yang2025cacheclip}, Cache-Craft \citep{agarwal2025cache}, and ProphetKV \citep{wang2026prophetkv} respectively use auxiliary models, historical information, or query-dependent signals.

% Source Word paragraph 16
In contrast, CacheTune constructs query-independent recomputation indices offline directly from each independent KV Chunk and uses them to guide KV transfer. Unlike FreqKV, which uses the frequency domain to compress KV storage \citep{kai2025freqkv}, CacheTune uses frequency-domain structure only to extract the positions of tokens important to restoring global cross-attention. Appendix~\ref{sec:A} discusses broader related work and background, with an illustrative comparison in Figure~\ref{fig:7}.

% Source Word paragraph 17

\section{Motivation}\label{sec:3}
% Source Word paragraph 18
\textbf{Restoring cross-chunk dependencies.} At layer $\ell$, the KV of token $i$ are derived from its hidden state, namely $k_i^{\ell}=h_i^{\ell}W_K^{\ell}$ and $v_i^{\ell}=h_i^{\ell}W_V^{\ell}$. This hidden state has already aggregated the preceding context through causal self-attention. Encoding a text chunk independently therefore omits its dependencies on preceding chunks, producing KVs different from those obtained by encoding the complete prompt.

% Source Word paragraph 19
We analyze KV Caches along the token dimension and observe that their energy is concentrated in low-frequency components. Appendix~\ref{sec:E.1} reports KV spectrum measurements for different models. This observation motivates us to use low-frequency structure to rank token positions for recomputation. Supplementary experiments in Appendix~\ref{sec:E.2} further show that low-frequency KV indices obtained through frequency-domain analysis provide high coverage of the KV positions that change most between independent chunk encoding and full-context encoding. 

\begin{figure}[!t]
\centering
\includegraphics[width=0.95\linewidth]{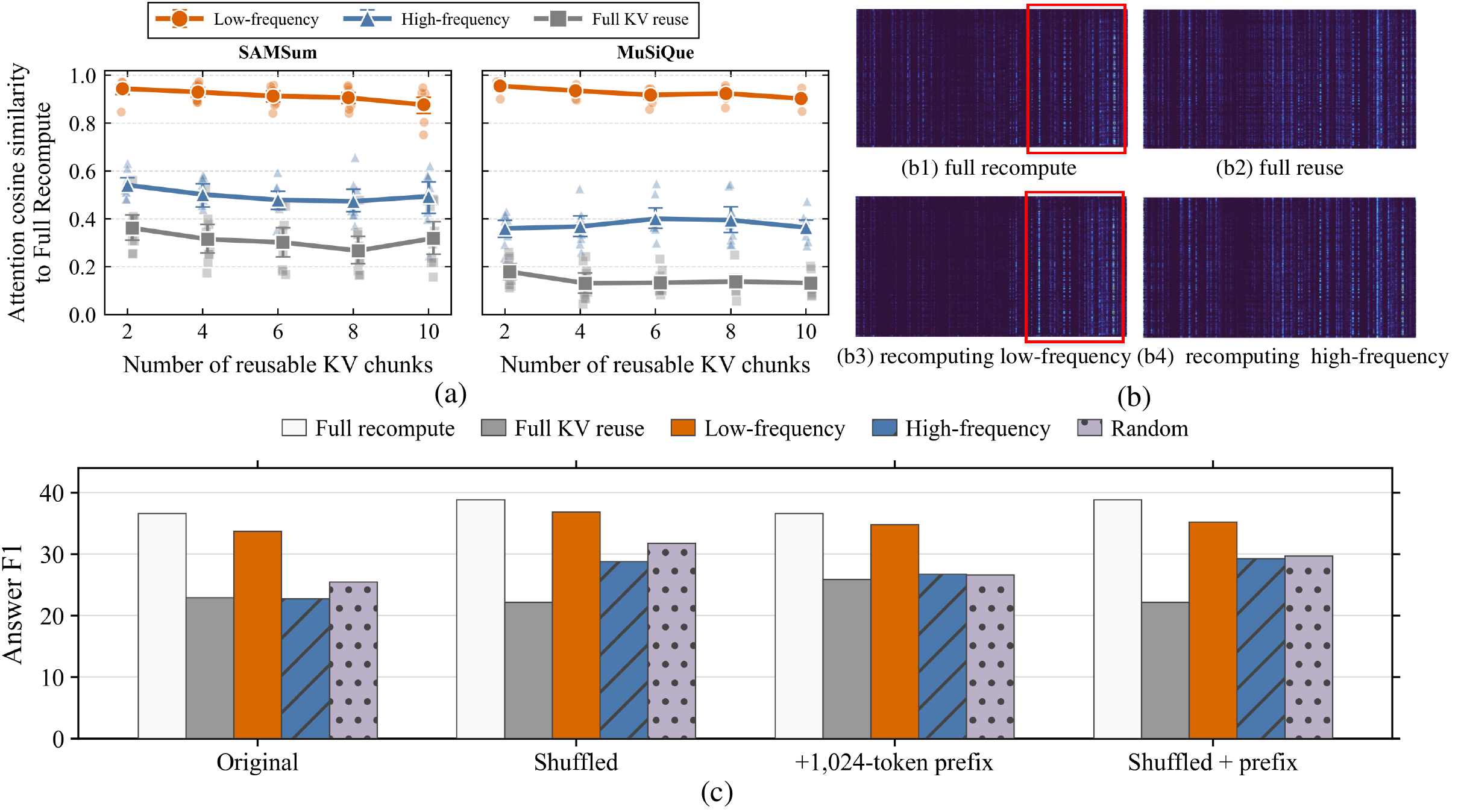}
\caption{Reusing low-frequency indices across contexts. (a) Attention similarity across different queries and numbers of chunks; curves and error bars show means and $95\%$ confidence intervals. (b) Representative attention distributions. (c) Mean F1 on MuSiQue with fixed caches and indices. All three selection strategies recompute $15\%$ of chunk tokens, and the added preceding context contains $1{,}024$ tokens. The model is Mistral v0.3-7B.}
\label{fig:1}
\end{figure}

% Source Word paragraph 20
To test whether this ranking helps restore missing dependencies, we compare low-frequency- and high-frequency-guided selection under the same $15\%$ recomputation budget, using full KV reuse and full recomputation as references. Full recomputation provides the exact full-context attention distribution. Figure~\ref{fig:1}(a,b) shows that low-frequency-guided recomputation more closely matches the attention distribution of full recomputation across different queries and numbers of chunks. Figure~\ref{fig:1}(c) further tests whether this improvement is reflected in the model's final accuracy. On MuSiQue, we fix the document chunk caches and indices while varying the query, chunk order, and length of the preceding context. Across all four context settings, low-frequency selection achieves higher F1 than full reuse, high-frequency selection, and random selection, approaching full recomputation. These results show that indices generated from chunk content can be reused across queries and context layouts without being selected again when a request arrives.

% Source Word paragraph 21
\textbf{Balancing computation and transfer.} Reusable KV Chunks may reside in CPU memory or on SSD/HDD. Increasing the recomputation ratio increases GPU workload but reduces the KV that must be loaded and allows more GPU computation to overlap with cache loading. In Appendix~\ref{sec:E.3}, we measure overall latency when different recomputation ratios are selected for KV Chunks across storage tiers. The results show that a higher reuse ratio is more suitable when cache access is inexpensive on fast storage media, whereas slower media may benefit from more recomputation. The selection of KV indices for recomputation must therefore be combined with complementary cache transfer, and the balance between them must be adjusted for the target storage tier.

% Source Word paragraph 25

\section{Methodology}\label{sec:4}
\begin{figure}[!t]
\centering
\includegraphics[width=.90\linewidth]{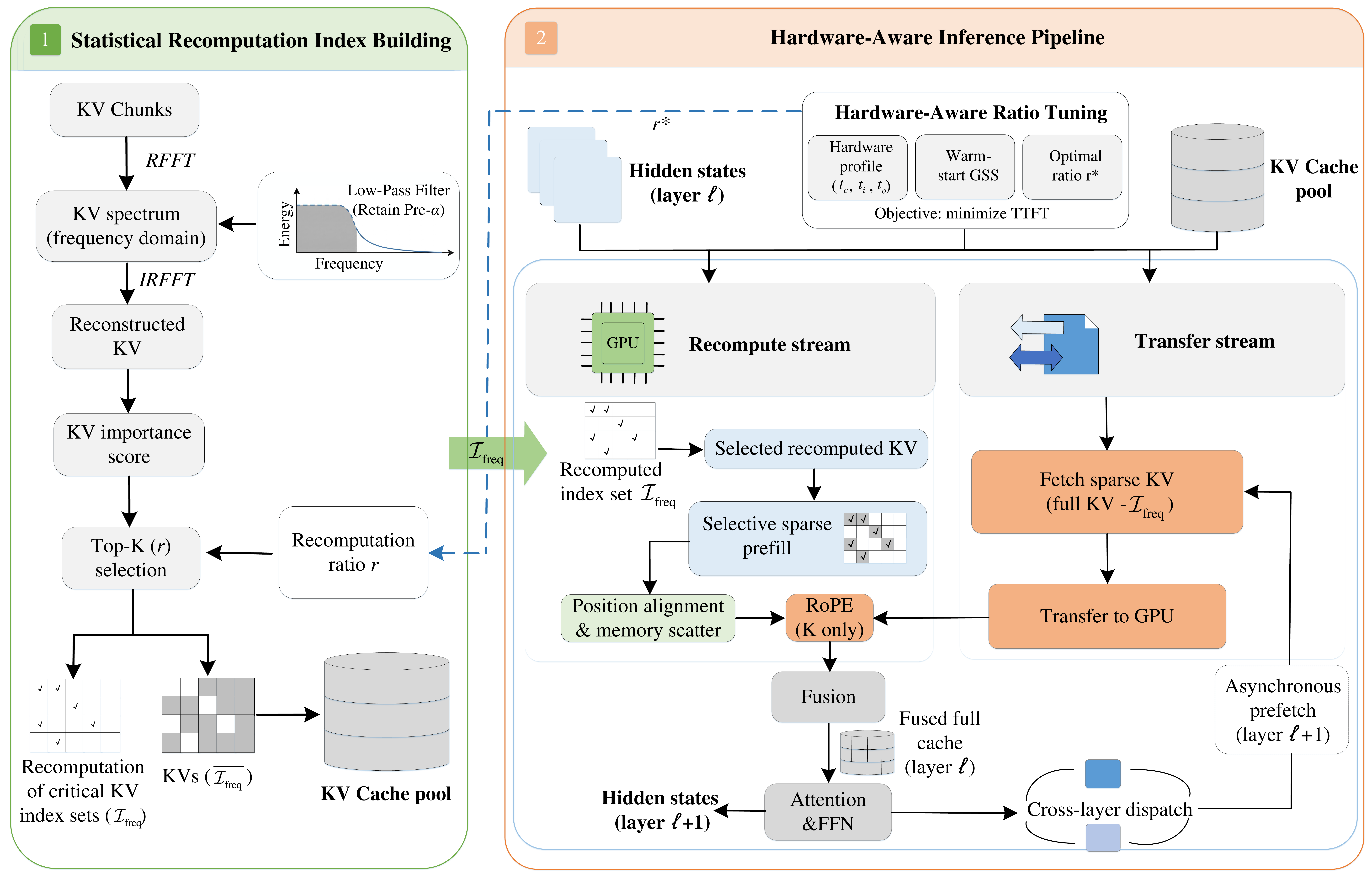}
\caption{Overall architecture of CacheTune.}
\label{fig:2}
\end{figure}

% Source Word paragraph 26
CacheTune operates in two stages, as shown in Figure~\ref{fig:2}. Offline frequency analysis ranks token positions in each independently encoded KV Chunk and determines the indices of positions that require recomputation to restore cross-attention. During online inference, KV at the selected positions is recomputed in the assembled context, while the complementary KV is transferred from the cache pool and merged with the recomputed results. The execution pipeline coordinates sparse transfer, selective recomputation, and positional recovery. A hardware-aware scheduler adjusts the recomputation ratio to balance these operations and reduce overall TTFT.

% Source Word paragraph 27
% Source Word paragraph 28
% Source Word paragraph 29

\subsection{Frequency-Domain Selective Recomputation}\label{sec:4.1}
% Source Word paragraph 30
During offline cache generation, CacheTune constructs a token-importance ranking for each independently encoded KV Chunk. When the chunk appears in a new context, this ranking is used to select the KV indices that need to be recomputed to restore cross-attention. The frequency-domain observations in Section~\ref{sec:3} provide the basis for this ranking. The procedure comprises spectral reconstruction, token scoring, and cross-layer selection.

% Source Word paragraph 31
\textbf{Recovering low-frequency structure aligned with tokens.} For layer $\ell$, let $\mathbf K^{(\ell)},\mathbf V^{(\ell)}\in\mathbb R^{N\times H\times D}$ denote the independently encoded Key and Value tensors, where $N$ is the chunk length, $H$ is the number of KV heads, and $D$ is the head dimension. We apply a real fast Fourier transform along the token axis separately for each head and feature channel \citep{williamson2012discrete}:
% Source Word paragraph 32
\begin{equation}
\widehat{\mathbf K}^{(\ell)}=\operatorname{rfft}(\mathbf K^{(\ell)},\mathrm{dim}=0),\qquad
\widehat{\mathbf V}^{(\ell)}=\operatorname{rfft}(\mathbf V^{(\ell)},\mathrm{dim}=0).
\label{eq:1}\end{equation}
% Source Word paragraph 33
Both spectra belong to $\mathbb C^{F\times H\times D}$, where $F=\lfloor N/2\rfloor+1$ contains the nonredundant frequencies of the real-valued sequence. Frequency indices describe variation across token positions; they are not themselves token indices.

% Source Word paragraph 34
We retain the lowest $\alpha$ fraction of frequencies and reconstruct a signal aligned with the original $N$ positions. Let $\mathbf X\in\{\mathbf K,\mathbf V\}$. The same operation applies to both Key and Value:
% Source Word paragraph 35
\begin{equation}\begin{aligned}
c&=\lfloor\alpha F\rfloor,\qquad
\widehat{\mathbf X}^{(\ell)}_{\mathrm{LP}}[k,:,:]=
\begin{cases}\widehat{\mathbf X}^{(\ell)}[k,:,:],&k<c,\\\mathbf 0,&k\ge c,\end{cases}\\
\widetilde{\mathbf X}^{(\ell)}&=\operatorname{irfft}(\widehat{\mathbf X}^{(\ell)}_{\mathrm{LP}},n=N,\mathrm{dim}=0).
\end{aligned}\label{eq:2}\end{equation}
% Source Word paragraph 36
Thus, the shapes of $\widetilde{\mathbf K}^{(\ell)}$ and $\widetilde{\mathbf V}^{(\ell)}$ return to $N\times H\times D$. This inverse transform allows us to assign importance scores to individual token positions based on the retained spectral structure. The cutoff ratio $\alpha$ controls which frequency components contribute to scoring, whereas the recomputation ratio $r$ controls how many token positions are updated. The former controls the basis for index selection, and the latter controls the online recomputation budget.

% Source Word paragraph 37
\textbf{Scoring and selecting positions shared across layers.} For token $i$, we compute the $\ell_2$ norm over its $H\times D$ elements to measure the magnitudes of the reconstructed Key and Value:
% Source Word paragraph 38
\begin{equation}\begin{aligned}
s_{i,K}^{(\ell)}&=\|\widetilde{\mathbf K}^{(\ell)}_i\|_2,\qquad
s_{i,V}^{(\ell)}=\|\widetilde{\mathbf V}^{(\ell)}_i\|_2,\\
s_i^{(\ell)}&=\tfrac12\bigl(s_{i,K}^{(\ell)}+s_{i,V}^{(\ell)}\bigr).
\end{aligned}\label{eq:3}\end{equation}
% Source Word paragraph 39
The joint score gives equal weight to KV magnitudes. Because numerical scales may differ across layers, we first normalize the scores within each layer and then average them over all $L$ layers:
% Source Word paragraph 40
\begin{equation}
\widetilde s_i^{(\ell)}=\frac{s_i^{(\ell)}}{\sum_{j=1}^N s_j^{(\ell)}+\epsilon},\qquad
\overline s_i=\frac1L\sum_{\ell=1}^L\widetilde s_i^{(\ell)}.
\label{eq:4}\end{equation}
% Source Word paragraph 41
Here, $\epsilon$ prevents division by zero. For recomputation ratio $r$, the shared index set and its complement are:
% Source Word paragraph 42
\begin{equation}
\mathcal I_{\mathrm{freq}}=\operatorname{TopK}(\{\overline s_i\}_{i=1}^N,rN),\qquad
\overline{\mathcal I}_{\mathrm{freq}}=\{1,\ldots,N\}\setminus\mathcal I_{\mathrm{freq}}.
\label{eq:5}\end{equation}
% Source Word paragraph 43
Here, TopK returns the token positions with the highest aggregated scores. Sharing these positions across layers allows updated hidden states to propagate continuously through the Transformer. Independently selecting unrelated positions in each layer would not form the same continuous recomputation path. The ranking depends only on the independently encoded text chunk and can therefore be reused directly by subsequent requests. Changing $r$ only changes the length of the prefix selected from this ranking; it does not require spectral analysis to be repeated.

% Source Word paragraph 44
\textbf{Recomputing in the new context.} During inference, local indices within each chunk are mapped to global positions in the assembled prompt. CacheTune propagates the hidden states at the selected positions forward layer by layer and regenerates their KV from the current layer's input. At each layer, this KV is merged at the original positions with the cached KV in the complement $\overline{\mathcal I}_{\mathrm{freq}}$. The selected queries therefore attend to the assembled preceding context under the causal mask, and their updated hidden states continue to the next layer. Both reused and recomputed Keys use the global positions described in Section~\ref{sec:4.2}.

% Source Word paragraph 45

\subsection{KV Cache Offloading and Sparse Reuse}\label{sec:4.2}
% Source Word paragraph 46
This section uses the offline indices obtained in Section~\ref{sec:4.1} to optimize non-prefix KV Cache reuse in two ways: reducing cache transfer volume and overlapping computation with transfer. These optimizations reduce the latency overhead of external cache access.

% Source Word paragraph 47
CacheTune stores reusable KV Chunks in cache pools outside the GPU, including CPU memory and SSD/HDD. The shared index set $\mathcal I_{\mathrm{freq}}$ guides both recomputation and cache loading: the system loads only the KV in the complement $\overline{\mathcal I}_{\mathrm{freq}}$, reducing the number of KV entries transferred per chunk per layer from $N$ to $(1-r)N$. Dedicated transfer and recomputation streams coordinate with the forward-computation stream through per-layer CUDA events, allowing asynchronous prefetching to overlap with computation. Reusable per-layer working buffers support scatter-based merging, placing transferred and recomputed KV back at their corresponding positions in the complete KV Cache.

% Source Word paragraph 48
To accommodate changes in chunk positions across requests, CacheTune stores Keys offline before rotary position embedding (RoPE) is applied. After reused Keys are transferred to the GPU, the recomputation stream applies RoPE according to their global positions in the assembled prompt; newly computed Keys use the same position mapping. Positional encoding is thus incorporated into the asynchronous pipeline and completed before attention uses the cache. Appendix~\ref{sec:B} and Figure~\ref{fig:8} further detail the dependencies between streams, buffer merging, and positional alignment.

% Source Word paragraph 49

\subsection{Hardware-Aware Joint Scheduling of Recomputation and Transfer}\label{sec:4.3}
% Source Word paragraph 50
The pipeline in Section~\ref{sec:4.2} overlaps recomputation with cache transfer, but the time taken by each path varies with the GPU and cache medium. A fixed recomputation ratio may leave a long wait for transfer on slower KV storage paths. Subject to an empirical lower bound for quality preservation, this section therefore treats the recomputation ratio as a joint decision variable for computation and transfer, aiming to shorten the critical path of the entire prefill pipeline.

% Source Word paragraph 51
The recomputation ratio $r$ trades GPU computational workload against cache transfer volume. Let $t_c$ and $t_i$ denote the per-token, per-layer computation cost and effective transfer cost obtained through profiling, respectively, and let $t_o$ denote the fixed overhead per layer. For $L$ layers and $N$ reusable tokens, the workloads of the recomputation and loading paths are proportional to $rN$ and $(1-r)N$, respectively. Approximating the main online prefill latency with a steady-state pipeline gives:
% Source Word paragraph 52
\begin{equation}
T_{\mathrm{TTFT}}(r)\approx L\max(rNt_c,(1-r)Nt_i)+Lt_o,\qquad
r_0=\frac{t_i}{t_c+t_i}.
\label{eq:6}\end{equation}
% Source Word paragraph 53
The maximum term corresponds to whichever of computation and transfer takes longer. Equating the times of the two paths yields the initial ratio $r_0$: the more expensive cache access is, the more beneficial it is to replace reads with recomputation. This approximation locates the initial search region; startup, synchronization, and pipeline bubbles in complete requests are accounted for by the measured objective below.

% Source Word paragraph 54
% Selecting $r$ based only on latency may result in too little recomputation for fast cache configurations. Based on the quality--latency results in Appendix~\ref{sec:E.4}, we set the lower search bound to $r_{\min}=15\%$ and select the recomputation ratio within $[r_{\min},r_{\max}]$, where $r_{\max}\leq 1$. For a small calibration set $\mathcal S_{\mathrm{cal}}$, we minimize the measured mean TTFT:
Selecting $r$ solely to minimize latency may leave too little
recomputation for fast cache configurations.
To limit generation-quality degradation, we set the lower search
bound to $r_{\min}=15\%$.
The quality--latency results in Appendix~\ref{sec:E.4} support this choice.
We search over $r\in[r_{\min},r_{\max}]$, where $r_{\max}\leq 1$,
and minimize the measured mean TTFT on a small calibration set
$\mathcal S_{\mathrm{cal}}$:
% Source Word paragraph 55
\begin{equation}
f(r)=\frac{1}{|\mathcal S_{\mathrm{cal}}|}\sum_{s\in\mathcal S_{\mathrm{cal}}}\operatorname{TTFT}(s;r),\qquad
r^*=\mathop{\arg\min}_{r\in[r_{\min},r_{\max}]}f(r).
\label{eq:7}\end{equation}
% Source Word paragraph 56
CacheTune clips $r_0$ to the search interval and uses this value as the initial point for a warm-started golden-section search \citep{kochenderfer2019algorithms}. Measuring TTFT in this way accounts for transfer startup, synchronization, and pipeline effects omitted from the theoretical cost model. The computation--transfer trade-off is approximately unimodal, providing a basis for this derivative-free search. Calibration is performed once when deploying to a given hardware/storage configuration, and the selected ratio can be used for subsequent requests. Appendix~\ref{sec:C} provides the detailed derivation and search pseudocode.

% Source Word paragraph 57

% Source Word paragraph 58

\section{Experimental Evaluation}\label{sec:5}
% Source Word paragraph 59

\subsection{Experimental Setup}\label{sec:5.1}
% Source Word paragraph 60
We use Mistral v0.3-7B, Llama 3-8B, and Qwen 2.5-32B \citep{DBLP:journals/corr/abs-2310-06825,DBLP:journals/corr/abs-2407-21783,qwen2024qwen25}. The datasets comprise SAMSum and Multi-News for summarization, evaluated using ROUGE-L, and MuSiQue, WikiMQA, and HotpotQA for multi-hop question answering, evaluated using F1. Experiments run on two platforms, with 2$\times$A100-40GB and 2$\times$RTX 4090-24GB GPUs, respectively; both connect to the host through PCIe Gen3$\times$16. Cache media include GPU memory, CPU memory, and HDD and SSD cache pools on the A100 and RTX 4090 platforms, respectively.

% Source Word paragraph 61
The baselines include Full Recompute, Full KV Reuse, vLLM Prefix Caching \citep{kwon2023efficient}, CacheBlend \citep{yao2025cacheblend}, EPIC \citep{DBLP:conf/icml/HuHWWHZFCS025}, KVLink \citep{DBLP:journals/corr/abs-2502-16002}, CacheSlide \citep{liu2026cacheslide}, and ProphetKV \citep{wang2026prophetkv}. Full Recompute processes the entire prompt; Full KV Reuse concatenates independently encoded chunk caches; Prefix Caching reuses only matching prefixes. CacheBlend uses a recomputation ratio of 15\% after selection in the first layer; EPIC recomputes the first 16 tokens of each chunk, and KVLink uses 5 link tokens per chunk. For CacheSlide, reusable documents are treated as fixed segments and queries as updated segments. ProphetKV aggregates query attention scores across layers and recomputes the top-scoring 15\% of context tokens. In the main CPU-cache comparison, CacheTune uses $\alpha=0.5$ and $r=15\%$; the storage-tier experiments also evaluate calibrated ratios. Appendix~\ref {sec:D} provides detailed baseline descriptions, dataset construction, and software versions.
%%measured storage bandwidth

% Source Word paragraph 62

\subsection{Generation Quality--TTFT Trade-off}\label{sec:5.2}
% Source Word paragraph 63
We compare CacheTune with the baselines on generation quality and TTFT. Methods such as EPIC, KVLink, and ProphetKV are not optimized for multi-tier KV Cache storage, so we keep all their reusable KV Chunks in GPU memory, the fastest-access tier, representing their best-case TTFT. CacheTune offloads reusable KV Chunks to CPU memory using the optimized pipeline in Section~\ref{sec:4.2}. With fast CPU-cache access, the hardware-aware scheduler selects the quality-preserving lower bound, $r=r_{\min}=15\%$. We set the low-frequency cutoff ratio to $\alpha=0.5$; Appendix~\ref{sec:E.4} evaluates sensitivity to both $r$ and $\alpha$.

% Source Word paragraph 64
% Source Word paragraph 65
\begin{figure}[!t]
\centering
\includegraphics[width=1.0\linewidth]{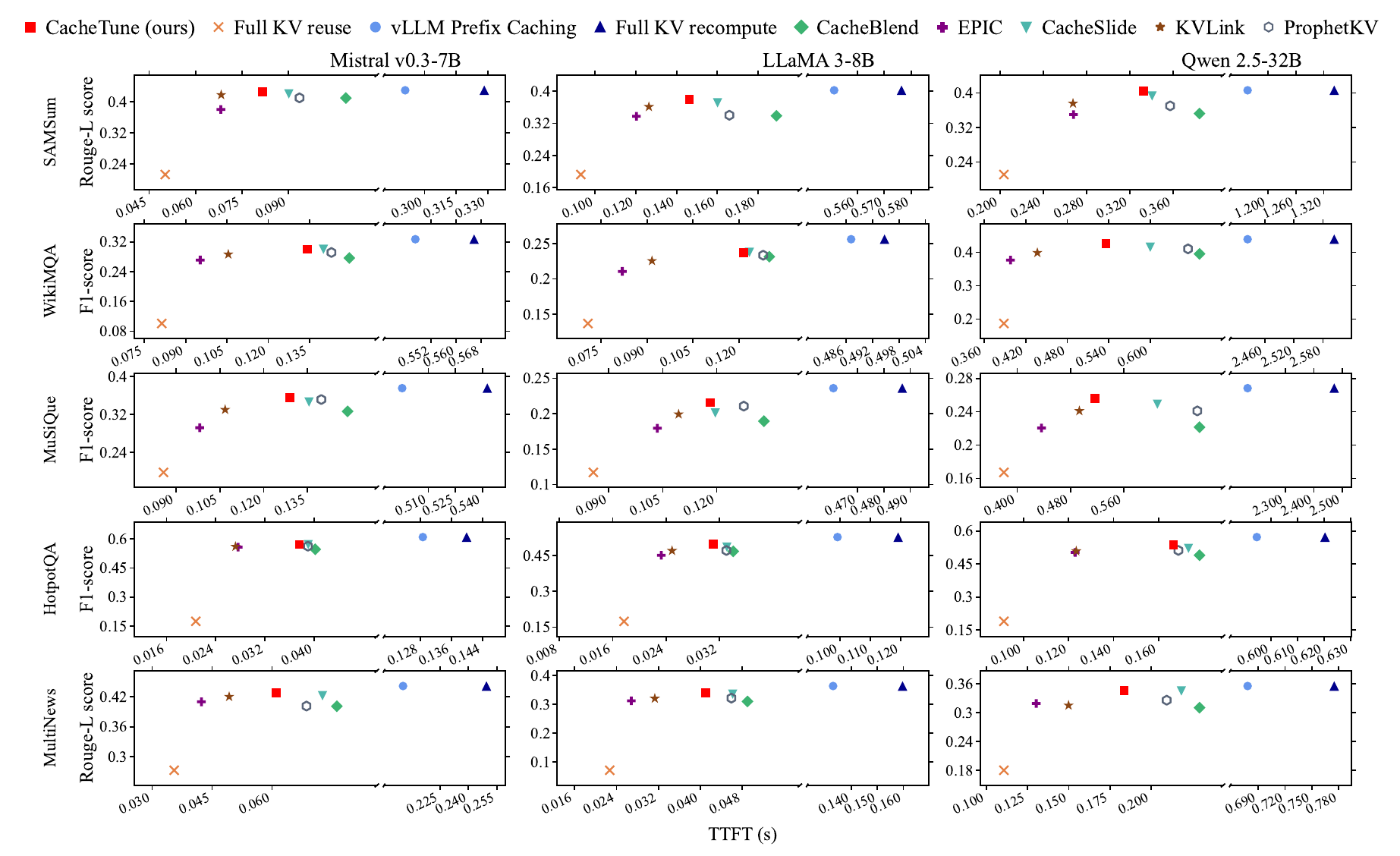}
\caption{Accuracy--TTFT trade-off of CacheTune and the baselines across models and dataset tasks.}
\label{fig:3}
\end{figure}
% Source Word paragraph 66
% Figure~\ref{fig:3} shows the overall trade-off. Full Recompute provides the accuracy upper bound but incurs the highest TTFT; Full KV Reuse has the lowest TTFT but severely reduces accuracy because cross-attention is missing. By losslessly reusing the KV Cache of matching prefixes, vLLM Prefix Caching preserves the accuracy of Full Recompute. However, it cannot exploit reusable non-prefix chunks that are dynamically combined in the workload. Many reusable chunks therefore still require recomputation, resulting in higher TTFT. CacheTune achieves accuracy close to Full Recompute, preserving an average of 94.8\% of its generation quality across five datasets and three models while delivering 3.72--4.86$\times$ TTFT speedup. At the same recomputation ratio of 15\%, CacheTune achieves both higher accuracy and lower TTFT than CacheBlend and ProphetKV. Its overall accuracy--TTFT trade-off also surpasses CacheSlide. CacheSlide primarily targets RPDC-style agent prompts, where reusable segments retain their relative order and positional drift is the main concern. CacheTune supports more general non-prefix reuse of independently encoded chunks with varying orders and positions. Together, these results show that CacheTune's frequency-domain selection restores cross-attention more effectively, while its sparse offloading and prefetching pipeline hides most transfer overhead.
Figure~\ref{fig:3} shows the overall trade-off. Full Recompute provides the accuracy upper bound but incurs the highest TTFT; Full KV Reuse has the lowest TTFT but severely reduces accuracy because cross-attention is missing. vLLM Prefix Caching preserves Full Recompute accuracy by losslessly reusing matching prefixes. However, many non-prefix chunks still require recomputation, resulting in higher TTFT than CacheTune. CacheTune achieves accuracy close to Full Recompute, preserving an average of 94.99\% of its generation quality across five datasets and three models while delivering 3.72--4.86$\times$ TTFT speedup. At the same recomputation ratio of 15\%, CacheTune achieves both higher accuracy and lower TTFT than CacheBlend and ProphetKV. Its overall accuracy--TTFT trade-off also surpasses CacheSlide. CacheSlide primarily targets RPDC-style agent prompts, where reusable segments retain their relative order and positional drift is the main concern. CacheTune supports more general non-prefix reuse of independently encoded chunks with varying orders and positions. Together, these results show that CacheTune's frequency-domain selection restores cross-attention more effectively, while its sparse offloading and prefetching pipeline hides most transfer overhead.

% Source Word paragraph 67
% CacheBlend and CacheSlide identify recomputation KVs through online first-layer recomputation, and ProphetKV selects recomputation targets by computing and aggregating query-to-context attention scores across layers online. CacheTune instead obtains recomputation indices offline during KV Chunk generation. Therefore, CacheTune can skip loading KV pairs that will be recomputed and transfer only the complementary KVs. This also avoids additional HKVD-style token-selection overhead. Very fast methods such as EPIC and KVLink achieve lower TTFT but suffer noticeable accuracy degradation. EPIC updates only a few fixed positions at the beginning of each KV Chunk, which may miss semantically important tokens in other regions and lead to accuracy loss across datasets. KVLink relies on pretrained link tokens, introducing additional offline computation. In contrast, CacheTune's frequency-domain analysis is lightweight: generating the recomputation index set for a 3K-token KV Chunk takes only about 0.69 ms on average on a GPU, and the indices can be reused across inference tasks.
CacheBlend and CacheSlide identify recomputation KVs through online first-layer recomputation, and ProphetKV selects recomputation targets by computing and aggregating query-to-context attention scores across layers online. CacheTune instead obtains recomputation indices offline during KV Chunk generation, avoiding online token-selection overhead. These indices also allow CacheTune to skip loading KV pairs that will be recomputed. Very fast methods such as EPIC and KVLink achieve lower TTFT but suffer noticeable accuracy degradation. EPIC updates only a few fixed positions at the beginning of each KV Chunk, which may miss semantically important tokens in other regions and lead to accuracy loss across datasets. KVLink relies on pretrained link tokens, introducing additional offline computation. In contrast, CacheTune's frequency-domain analysis is lightweight: generating the recomputation index set for a 3K-token KV Chunk takes only about 0.69 ms on average on a GPU, and the indices can be reused across inference tasks.
% Source Word paragraph 68

\subsection{Inference Serving Throughput}\label{sec:5.3}
\begin{figure}[!t]
\centering
\begin{minipage}[t]{.55\linewidth}
\vspace{0pt}
\centering
\includegraphics[width=\linewidth]{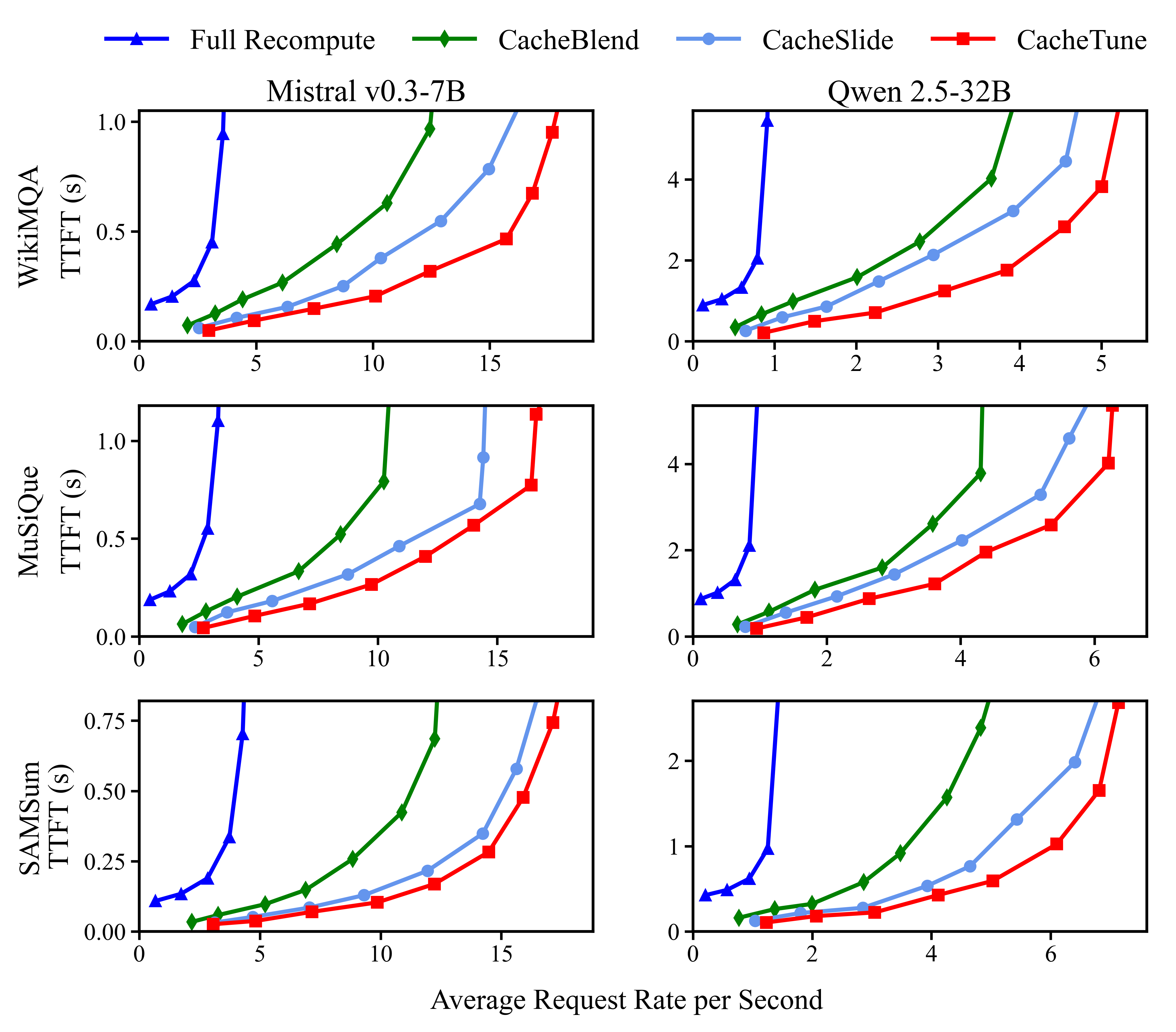}
\caption{TTFT as the request rate increases. Curves extending farther to the right at low TTFT indicate higher effective throughput under low-latency serving conditions.}
\label{fig:4}
\end{minipage}\hfill
\begin{minipage}[t]{.425\linewidth}
\vspace{0pt}
\captionof{table}{TTFT (s) comparison between GPU-resident reuse and the optimized CPU offloading pipeline.}
\label{tab:1}
\centering
\footnotesize
\setlength{\tabcolsep}{1.8pt}
\renewcommand{\arraystretch}{1.06}
\begin{tabular}{@{}lrrr@{}}
\toprule
\multicolumn{4}{c}{GPU} \\
\midrule
Method & SAMSum & WikiMQA & MuSiQue \\
\midrule
\shortstack[l]{Full\\Recompute} & 0.3283 & 0.5657 & 0.5426 \\
\shortstack[l]{Prefix\\Caching} & 0.2908 & 0.5469 & 0.4955 \\
CacheBlend & 0.1086 & 0.1495 & 0.1487 \\
CacheSlide & 0.0902 & 0.1401 & 0.1355 \\
CacheTune & 0.0899 & 0.1485 & 0.1398 \\
\midrule
\multicolumn{4}{c}{CPU} \\
\midrule
% Method & SAMSum & WikiMQA & MuSiQue \\
% \midrule
\shortstack[l]{Full\\Recompute} & 0.3283 & 0.5657 & 0.5426 \\
\shortstack[l]{Prefix\\Caching} & 0.3012 & 0.5496 & 0.5071 \\
CacheBlend & 0.1113 & 0.1768 & 0.1723 \\
CacheSlide & 0.0995 & 0.1404 & 0.1492 \\
CacheTune & 0.0817 & 0.1342 & 0.1289 \\
\bottomrule
\end{tabular}
\end{minipage}
\end{figure}

% Source Word paragraph 69
As shown in Figure~\ref{fig:4}, we compare the TTFT of different methods at varying request rates. Full Recompute saturates at relatively low request rates, causing TTFT to rise rapidly. CacheBlend and CacheSlide improve throughput, but their latency still increases considerably under heavy load. CacheTune maintains low TTFT while shifting saturation to higher request rates, thereby achieving higher throughput. Relative to Full Recompute, CacheTune achieves 3.93--6.21$\times$ throughput, compared with 2.71--5.16$\times$ for CacheBlend and 3.79--5.94$\times$ for CacheSlide. This shows that CacheTune reduces per-request prefill latency and improves effective serving capacity across load levels.

% Source Word paragraph 70
% Source Word paragraph 71
% Source Word paragraph 72

\subsection{Effectiveness of Frequency-Guided Selection}\label{sec:5.4}
\begin{figure}[!tb]
\centering
\begin{minipage}[t]{.49\linewidth}
\vspace{0pt}
\centering
\includegraphics[width=\linewidth]{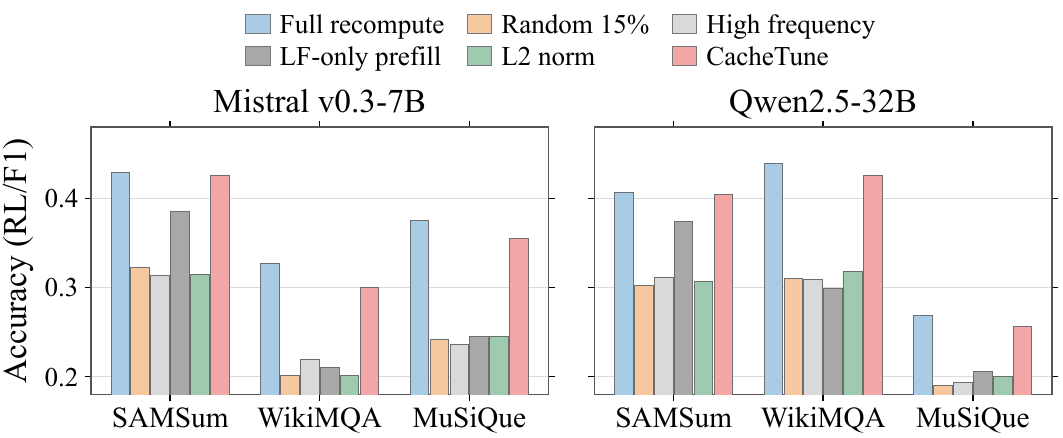}
\caption{Effectiveness of different strategies for selecting tokens for recomputation.}
\label{fig:5}
\end{minipage}\hfill
\begin{minipage}[t]{.49\linewidth}
\vspace{0pt}
\centering
\includegraphics[width=\linewidth]{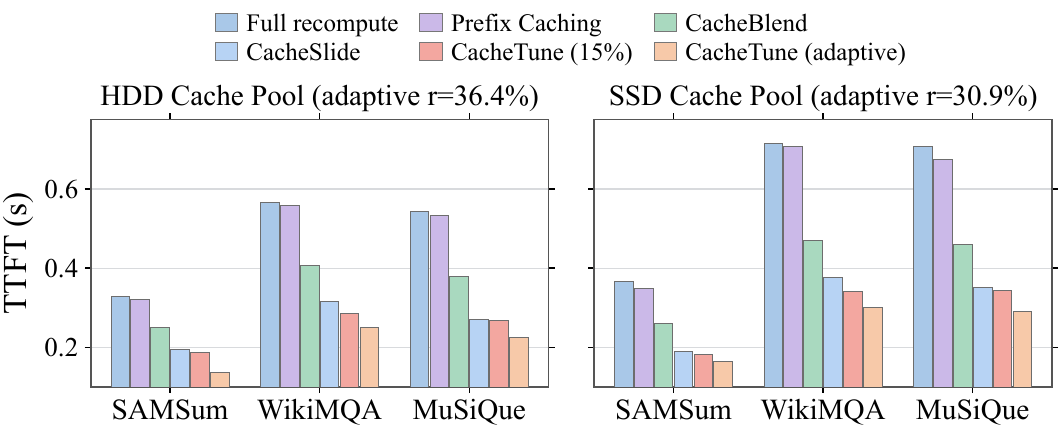}
\caption{TTFT with different external KV Cache media and recomputation strategies.}
\label{fig:6}
\end{minipage}
\end{figure}

To assess whether frequency-domain importance scores identify the KV pairs most critical to restoring global cross-attention, we compare five strategies at a fixed 15\% token-selection ratio: random selection, L2-norm selection, high-frequency-guided selection, LF-only prefill, and CacheTune. LF-only prefill retains the top 15\% of tokens ranked by low-frequency importance and discards the rest before prefill. L2-norm selection uses the same joint K/V norm and cross-layer aggregation as in Section~\ref{sec:4.1}, but omits frequency-domain analysis. As shown in Figure~\ref{fig:5}, random, L2-norm, and high-frequency-guided selection recover only a limited portion of the quality lost through non-prefix KV reuse and do not effectively restore cross-chunk semantic dependencies. LF-only prefill performs relatively well on SAMSum, where summarization mainly requires understanding the overall contextual semantics, suggesting that low-frequency analysis can identify tokens with strong semantic representativeness. However, its performance drops markedly on WikiMQA and MuSiQue, which require fine-grained evidence and multi-hop reasoning, and remains below that of CacheTune across all settings in Figure~\ref{fig:5}. These results show that tokens selected by low-frequency analysis help restore cross-chunk attention but cannot alone form a complete context meeting task requirements. The discarded tokens still contain important factual details, local semantics, and intermediate reasoning evidence. CacheTune consistently produces results closest to Full Recompute, showing that recomputing the identified critical KV pairs in the global context while reusing the remaining KV pairs helps restore cross-chunk dependencies without discarding the rest of the context.

% Source Word paragraph 74
% Source Word paragraph 75
% \begin{figure}[!t]
% \centering
% \includegraphics[width=.90\linewidth]{figures/figure05.png}
% \caption{Effectiveness of different strategies for selecting tokens for recomputation.}
% \label{fig:5}
% \end{figure}
% Source Word paragraph 76
\subsection{Adaptability Across Cache Tiers}\label{sec:5.5}
% Source Word paragraph 77
\textbf{GPU and CPU caches.} This experiment examines whether the external-cache pipeline preserves latency gains after moving the cache out of GPU memory. Table~\ref{tab:1} compares Mistral with GPU-resident and CPU-resident caches, with CacheTune's recomputation ratio fixed at 15\%. Within each comparison, CacheBlend, CacheSlide, and Prefix Caching use the same cache medium. With CPU-resident caches, CacheTune reduces average TTFT by 25.3\%, 11.9\%, and 74.4\% relative to these three baselines, respectively. Across the three tasks, its CPU-cache TTFT is comparable to or lower than that of its own GPU-resident reuse path. This is because the CPU configuration enables the pipeline that overlaps transfer and recomputation, while sparse loading and deferred RoPE further reduce GPU-side data processing and computation. These end-to-end measurements show that the optimized CPU-cache path preserves the benefits of reuse despite the additional data movement.

% Source Word paragraph 78
% Table 1 is paired with Figure 4 above; values are unchanged.
% Source Word paragraph 121
\textbf{SSD/HDD caches.} Using Mistral, we further compare TTFT with a fixed $r=15\%$ and with hardware-aware calibration on HDD and SSD configurations to examine the effect of adjusting the recomputation budget. In Figure~\ref{fig:6}, the fixed $r=15\%$ achieves 1.92$\times$ and 2.05$\times$ TTFT speedup over Full Recompute on the HDD and SSD configurations, respectively. Calibration selects $r=36.4\%$ for the HDD and $r=30.9\%$ for the SSD, increasing the respective speedups to 2.36$\times$ and 2.34$\times$. More recomputation reduces costly cache reads, further lowering end-to-end latency. These ratios are calibrated for latency using the first 10 SAMSum samples. Calibration takes approximately 3--4 minutes for each hardware and storage configuration, after which the ratios are reused across datasets without recalibration. This calibration estimates the balance between computation and I/O, while the lower bound on $r$ prevents severe generation-quality degradation from insufficient recomputation. Appendix~\ref{sec:E.5} provides further analysis of these results and quantifies the latency benefits of the cache-loading optimizations introduced in Section~\ref{sec:4.2} through cumulative ablation.

% Source Word paragraph 122
% Source Word paragraph 123
% \begin{figure}[!t]
% \centering
% \includegraphics[width=.90\linewidth]{figures/figure06.png}
% \caption{TTFT with different external KV Cache media and recomputation strategies.}
% \label{fig:6}
% \end{figure}
% Source Word paragraph 124

Appendix~\ref{sec:F} further discusses the experimental results and the method’s limitations.
% \subsection{Limitations}\label{sec:5.6}
% % Source Word paragraph 125
% CacheTune may still be I/O-bound when external storage bandwidth is extremely low or asynchronous overlap is insufficient. Frequency-guided ranking is an empirical signal that has been evaluated on the models and tasks above. Higher concurrency, more complex workloads, and CXL-based disaggregated memory \citep{10.1145/3669900,seo2025oasis} remain to be evaluated.

% Source Word paragraph 126
\Needspace{5\baselineskip}

\section{Conclusion}\label{sec:6}
% Source Word paragraph 127
We present CacheTune to address cross-chunk attention loss and cross-tier I/O bottlenecks in non-prefix KV Cache reuse for long-context LLM serving. CacheTune uses the low-frequency structure of independently encoded KV Chunks to identify the KV pairs most critical to restoring global semantics. It coordinates sparse transfer, selective recomputation, deferred RoPE restoration, and hardware-aware scheduling across GPU, CPU, and SSD/HDD cache tiers. Across multiple models and long-context tasks, CacheTune preserves generation quality close to Full Recompute while achieving 3.72--4.86$\times$ TTFT speedup and 3.93--6.21$\times$ inference serving throughput.

\FloatBarrier
\label{main:end}
\bibliography{reference}
\bibliographystyle{cachetune_references}
\appendix
% Source Word paragraph 170

\section{Supplementary Background on Cache Reuse and Frequency-Domain Methods}\label{sec:A}
% Source Word paragraph 171
As context lengths and serving scales grow, the KV Cache is evolving from temporary state within a single inference run into data that can be retained, retrieved, and migrated across requests. Tiered caching studies how to use the capacities and access speeds of different media \citep{gao2024cost,jin2025ragcache}, global scheduling coordinates requests and cache resources \citep{qin2024mooncake}, and vector indexing and multi-tier data placement support cache organization and access at larger scales \citep{10.14778/3796195.3796212,lin2026kvdrive}. On top of these management mechanisms, reusing existing KVs can eliminate prefill computation for repeated text. How this reuse is implemented depends on whether the context in which the cache was generated matches that of the current request.

% Source Word paragraph 172
Prefix caching uses this consistency of context: when two requests share the same prefix, the preceding context of each token in that prefix remains unchanged, allowing the corresponding KVs to be reused directly. vLLM, SGLang, and ChunkAttention all use prefix sharing to reduce repeated computation \citep{kwon2023efficient,zheng2024sglang,ye2024chunkattention}. In RAG and multi-document summarization, however, reusable content often consists of independent document chunks whose order and positions vary across requests. When independently encoded chunks enter a new prompt, they need to establish dependencies on preceding chunks, giving rise to the problem of cross-chunk repair in non-prefix cache reuse.

% Source Word paragraph 173
Non-prefix reuse methods adopt different organization and repair mechanisms. PromptCache uses predefined prompt schemas to organize KVs into modules that can be cached and composed independently \citep{gim2024prompt}. To address the loss of cross-attention after independent chunks are concatenated, CacheBlend fully recomputes the first layer at runtime and uses KV deviations to select tokens to update in subsequent layers \citep{yao2025cacheblend}; EPIC concentrates recomputation at attention-sink positions at the beginning of each chunk \citep{DBLP:conf/icml/HuHWWHZFCS025}; and KVLink learns link tokens at chunk boundaries through additional training to establish connections between chunks \citep{DBLP:journals/corr/abs-2502-16002}. These designs organize repair using numerical changes in the current context, fixed positional priors, and learned connection mechanisms, respectively. Prior methods also adjust the recomputation ratio $r$ based on estimated computation and cache-loading costs \citep{yao2025cacheblend}. CacheTune further considers how $r$ affects complementary transfer volume and calibrates the ratio using measured end-to-end TTFT, accounting for factors not captured by these estimates, such as pipeline startup, synchronization, and overlap effects, to achieve more precise adjustment of the recomputation ratio.

% Source Word paragraph 174
Other methods organize updates around the reuse scenario or selection signal. CacheSlide studies relative-position-dependent caching, where reusable segments retain their relative order while their absolute positions shift, and supports reuse through chunked contextual positional encoding, weighted correction attention, and cache management optimizations \citep{liu2026cacheslide}. CacheClip uses a fine-tuned auxiliary model to guide token selection \citep{yang2025cacheclip}, Cache-Craft estimates chunk reusability from prefix overlap and historical attention \citep{agarwal2025cache}, and ProphetKV combines query-induced attention with multi-layer metrics to prioritize recomputation \citep{wang2026prophetkv}. These studies expand the structural, historical, and query information available for cache reuse.

\begin{figure}[tbp]
\centering
\includegraphics[width=1.0\linewidth]{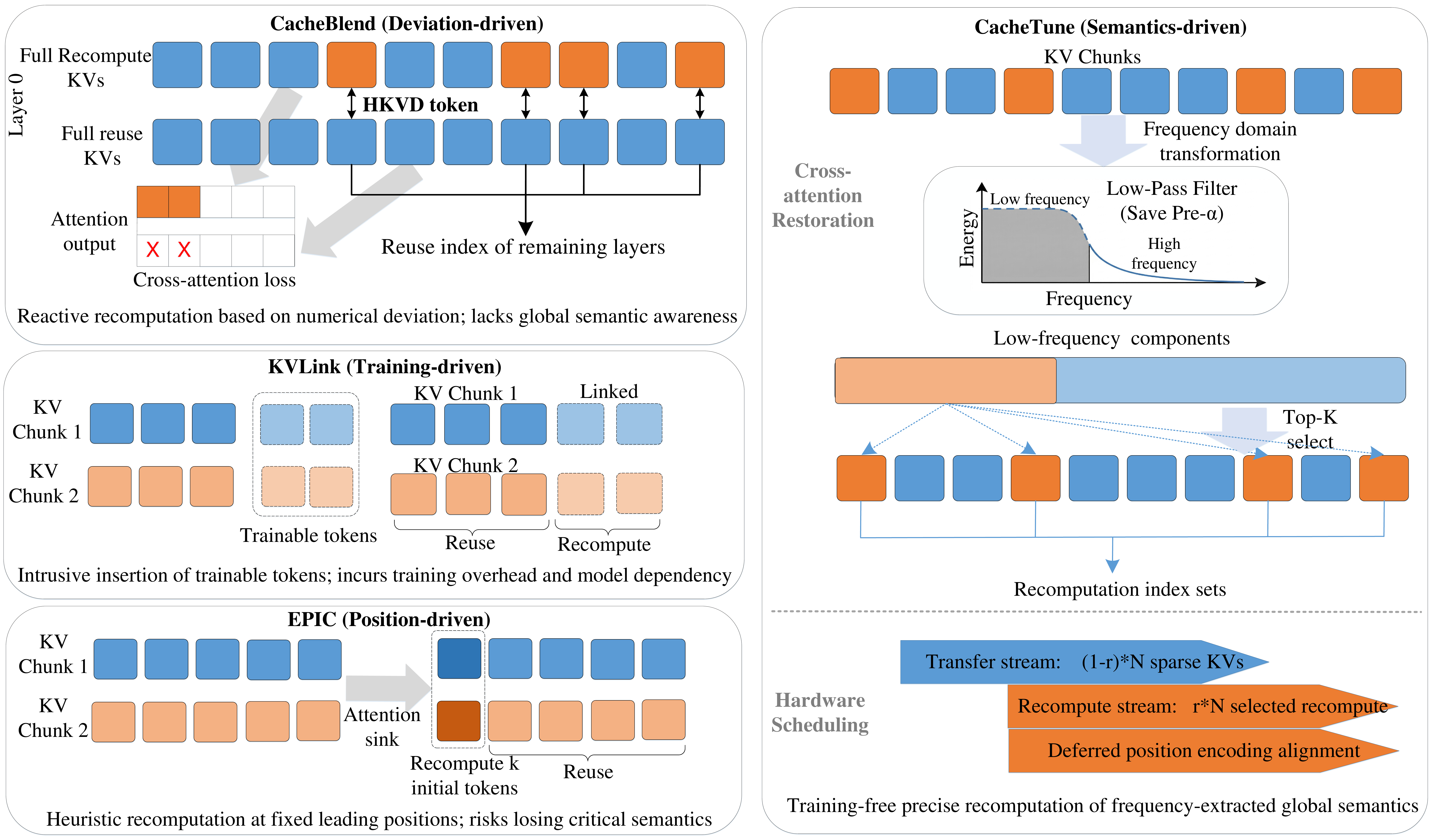}
\caption{Comparison of cross-attention recovery strategies for non-prefix KV Cache reuse.}
\label{fig:7}
\end{figure}

% Source Word paragraph 175
CacheTune extracts a ranking of positions from the KV frequency-domain structure of independently encoded chunks, constructing the basis for selection during cache generation. When a request arrives, it recomputes a subset of tokens according to this ranking so that their representations incorporate the current global context, and loads cached KVs for the remaining positions. Because the ranking is generated from the chunk itself, the same chunk can retain its existing indices across different queries, and the cache system can use them to determine in advance which complementary portion to transfer. Figure~\ref{fig:7} contrasts the positions selected and updated within a chunk by several cross-chunk repair strategies, illustrating the differences in their selection criteria and recomputation schemes.

% Source Word paragraph 176
% Source Word paragraph 177
% Source Word paragraph 178
In image and signal processing, different frequency components describe variations at different scales, and their energy distribution can guide representation and compression \citep{xu2020learning,wang2023division,liu2026fdsr}. FreqKV applies this idea to the KV Cache, reducing storage overhead by retaining low-frequency structure and compressing high-frequency components \citep{kai2025freqkv}. CacheTune instead uses low-frequency structure to identify recomputation positions that are important for recovering cross-attention. The KVs at every original position are supplied through either recomputation or reuse, preserving the full context while concentrating the limited recomputation budget on repairing important cross-chunk cross-attention.

% Source Word paragraph 179

\section{Offline Preparation and Runtime Implementation}\label{sec:B}
\begin{figure}[tbp]
\centering
\includegraphics[width=1.0\linewidth]{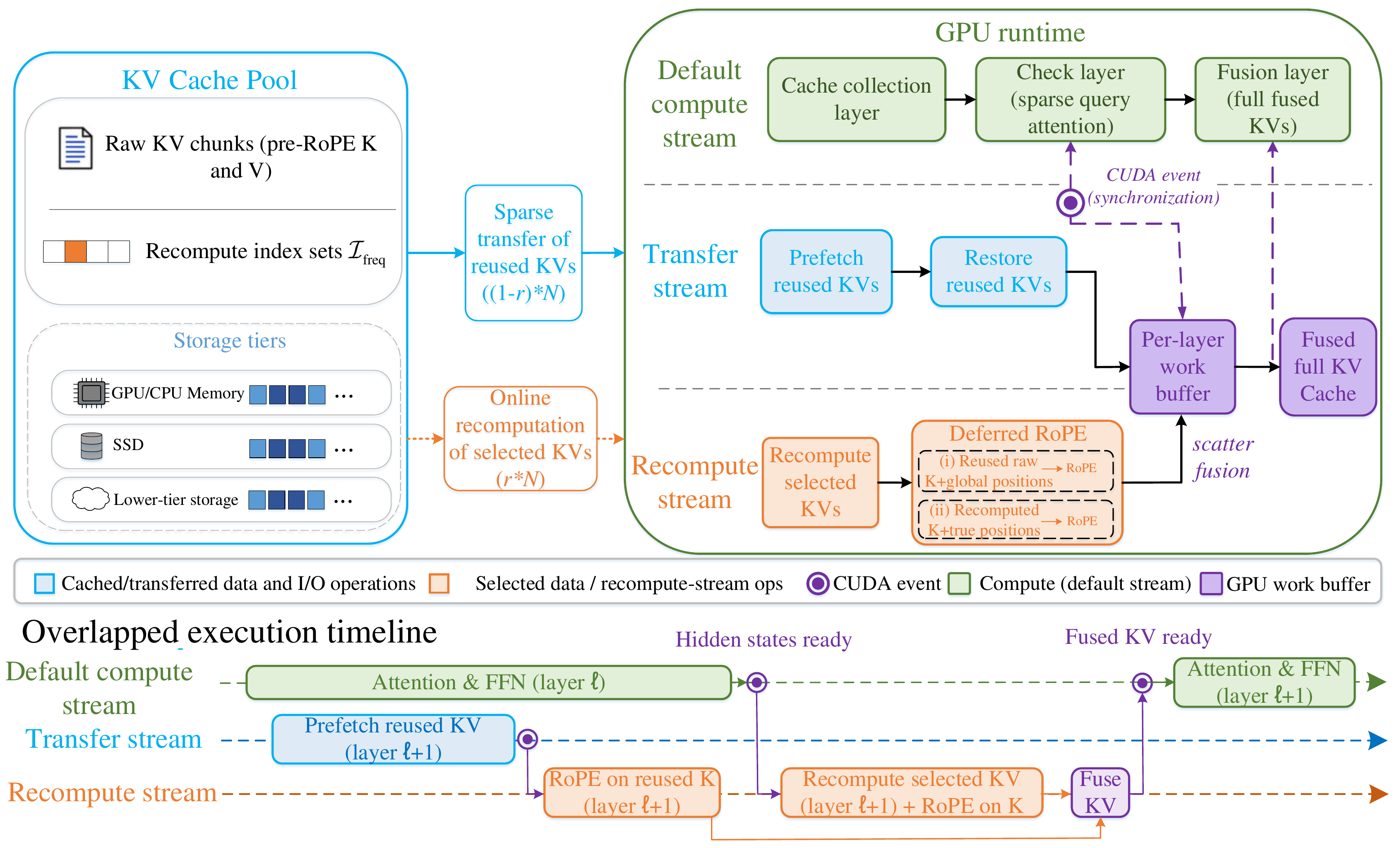}
\caption{KV Cache offloading and sparse reuse pipeline.}
\label{fig:8}
\end{figure}

% Source Word paragraph 180
CacheTune uses independently encoded KV Chunks together with frequency-guided selection of important KV indices for online prefill. The important KV indices determine which tokens to update at each layer, the cache pool supplies KVs for the remaining positions, and the runtime fuses the two parts according to their positions in the composed prompt. The same index set thus connects the selection rule in Section~\ref{sec:4.1} with the execution pipeline in Section~\ref{sec:4.2}: once a position is selected, its KVs are generated by forward computation in the current context; the remaining positions follow the sparse loading path. Figure~\ref{fig:8} shows the layer-by-layer execution of this process.
% Source Word paragraph 181

\subsection{Offline Preparation}\label{sec:B.1}
% Source Word paragraph 182
For a reusable text chunk containing $N$ tokens, CacheTune first encodes it independently and collects the Keys and Values from all $L$ layers. The tensor at each layer has shape $N\times H\times D$, where $H$ is the number of KV heads and $D$ is the head dimension. Spectral analysis is performed along the token axis, so the KV representations that vary across positions in each head and channel form a sequence of length $N$. Because the sequences are real-valued, rFFT produces $\lfloor N/2\rfloor+1$ non-redundant frequencies. CacheTune retains the lowest $\alpha$ fraction of frequency components and applies the inverse transform with the output length set to $N$ to recover low-frequency tensors aligned with the original positions.

% Source Word paragraph 183
Following the importance scoring rule in Section~\ref{sec:4.1}, CacheTune computes the magnitudes of the low-frequency Key and Value at each position separately, combines them with equal weights, normalizes the scores within each layer, and averages them across layers to form a unified importance ranking within the chunk. Normalization within each layer makes the numerical scales of different layers comparable, while cross-layer aggregation selects the same set of positions for continuous sparse forward execution. For a given recomputation ratio $r$, the top-ranked $rN$ positions form the recomputation set $\mathcal I_{\mathrm{freq}}$, and its complement is denoted by $\overline{\mathcal I}_{\mathrm{freq}}$. The hidden states of the former propagate continuously across Transformer layers, while the latter supply context through cached KVs at each layer.

% Source Word paragraph 184
The ranking is computed during cache generation, and subsequent requests select the corresponding number of positions according to $r$. The spectral transform for a single head and channel has a complexity of $O(N\log N)$. Accounting for the number of layers, KV heads, and head dimension gives a total complexity of $O(LHDN\log N)$ for this step, while scoring and aggregation require $O(LHDN)$. This preprocessing overhead is amortized over repeated uses of a cached chunk. Section~\ref{sec:5.2} reports that generating recomputation indices on the GPU for a cached chunk containing 3K tokens takes only about 0.69 ms on average.

% Source Word paragraph 185
Once a new composed prompt is formed, CacheTune maps local indices within each chunk to global indices using the chunk's starting position in the sequence. If chunk $b$ starts at global position $o_b$, its local position $i$ corresponds to $o_b+i-1$. The recomputation set and its reuse complement use the same mapping for query filtering, cache loading, fusion writeback, and positional encoding. Changing the order or offsets of chunks therefore requires only an update to the index mapping, while adjusting the recomputation ratio changes the set size along the stored ranking.

% Source Word paragraph 186

\subsection{Sparse KV Reuse Pipeline}\label{sec:B.2}
% Source Word paragraph 187
Online prefill assembles KVs from two sources at each layer: KVs in $\mathcal I_{\mathrm{freq}}$ are recomputed in the current context, and KVs in $\overline{\mathcal I}_{\mathrm{freq}}$ are loaded from the external cache pool. CacheTune applies this partition at the start of loading, reducing the transfer volume for each chunk at each layer from $N$ to $(1-r)N$ KV pairs. Online computation retains the state-update path of the selected tokens, while the KVs accessed by attention cover the full sequence. To connect these two granularities, the runtime uses a collection layer, a check layer, and subsequent fusion layers, with reusable per-layer working buffers for KV assembly.

% Source Word paragraph 188
The collection layer organizes the offline KV Chunks to be reused by the current request and establishes the positional correspondence between these chunks and the composed prompt. The precomputed ranking of each chunk is thereby converted into request-level recomputation and loading indices. Cached chunks may reside in CPU memory, SSDs, or HDDs. The runtime uses the loading indices to obtain KVs in the reuse complement at each layer, leaving the selected positions to be generated by the recomputation path. Chunk content, layer number, and global position are associated at this stage for subsequent layer-by-layer use.

% Source Word paragraph 189
The check layer applies the precomputed indices on the query side to select tokens in reusable chunks that require further updates. Queries at the selected positions participate in the current layer's attention computation and aggregate preceding context over the complete KVs of that layer. After attention and the remaining forward computation, the hidden states at these positions pass to the next layer to continue generating the required queries and KVs. The same index set is retained across layers so that every selected position has a continuously updated hidden state; uncached content, such as new queries, still follows the normal forward path.

% Source Word paragraph 190
Layers after the check layer perform fusion. At layer $\ell$, the recomputation path generates KVs in $\mathcal I_{\mathrm{freq}}$ from the updated input hidden states, while the transfer path obtains the corresponding layer's cached KVs in $\overline{\mathcal I}_{\mathrm{freq}}$. After global positional encoding is applied to both sets of Keys, the system writes reused KVs into the working buffer according to their position indices, then scatters recomputed KVs into the remaining selected positions. These two complementary index sets cover all positions in the reusable chunks at that layer and, together with KVs generated for uncached content such as new queries, form the complete cache. The selected queries of the current layer access the preceding context under a causal mask, and the results undergo operations within the layer, such as feed-forward computation, to form the input to the next layer. Reusable buffers allow this assembly process to proceed layer by layer while reducing repeated allocation and intermediate data organization overhead.

% Source Word paragraph 191
These operations are coordinated across three CUDA streams. The model's native forward stream performs the current layer's forward computation; the transfer stream asynchronously prefetches the reused KVs needed by the next layer; and the recomputation stream generates selected KVs from ready hidden states and performs positional recovery for reused Keys. The streams hand off data through per-layer CUDA events: after transfer completes, reused Keys undergo positional recovery; once both the recomputed and reused portions are ready, the fused result is consumed by the current layer's attention; and the layer's output then supplies the input for recomputation at the next layer.

% Source Word paragraph 192
The pipeline advances continuously through the layers. While the forward stream processes layer $\ell$, the transfer stream prefetches the reuse complement for layer $\ell+1$, overlapping external cache access with the current layer's computation. Once the output of layer $\ell$ is ready, the recomputation stream can use it to generate the selected KVs for the next layer and combine them with the prefetched result to assemble that layer's cache. Per-layer events schedule handoffs at the points where data are consumed, while prefetching uses computation time in adjacent layers to complete loading in advance. The reduction in transfer volume enabled by offline indices thus works together with cross-layer overlap to shorten the time online prefill spends waiting for external cache loading.

% Source Word paragraph 193
% Source Word paragraph 194

% Source Word paragraph 195

\subsection{Deferred RoPE for Positional Alignment}\label{sec:B.3}
% Source Word paragraph 196
The local positions within an independent chunk usually differ from its global positions in the new context. In models that use RoPE, positional rotation is applied to Keys after linear projection. If cached Keys have already been rotated according to their positions within the chunk, moving the chunk directly carries the original positional encoding into the new sequence. CacheTune therefore stores raw pre-RoPE Keys in the offline cache and defers positional rotation until online reuse, allowing the same chunk to be used at different global offsets.

% Source Word paragraph 197
After the transfer stream loads the raw reused Keys of layer $\ell$ onto the GPU, the recomputation stream reads their corresponding global positions $p_i^{\mathrm{global}}$ and performs positional recovery as follows:
% Source Word paragraph 198
\begin{equation}
\mathbf K_{\mathrm{reuse},i}^{(\ell)}=\operatorname{RoPE}(\mathbf K_{\mathrm{raw},i}^{(\ell)},p_i^{\mathrm{global}}),\qquad
i\in\overline{\mathcal I}_{\mathrm{freq}}.
\label{eq:8}\end{equation}
% Source Word paragraph 199
Newly generated Keys at selected positions likewise receive RoPE using their true global positions, while Values participate directly in fusion according to their corresponding indices. Both types of Keys are thus expressed in the same positional coordinate system of the composed prompt, allowing queries at the current layer to perform causal attention over the fused KVs. Positional recovery aligns the chunk's coordinates within the new prompt, while selective recomputation restores cross-chunk dependencies by updating hidden states; the two operate together within the same fusion process.

% Source Word paragraph 200
Deferred RoPE uses the same index mapping as sparse loading and is integrated into the asynchronous path of the recomputation stream. Reused Keys are handed off in the order of transfer completion, positional recovery, and consumption during fusion, interleaved with ready computation or transfer tasks from other layers. Positional encoding is thus recovered along the online data flow, while chunks in the cache pool retain a uniform form that can be reused across requests.

% Source Word paragraph 201

\section{Hardware-Aware Calibration of the Recomputation Ratio}\label{sec:C}
% Source Word paragraph 202
The complementary construction in Section~\ref{sec:4.2} makes the recomputation ratio $r$ control two costs simultaneously: increasing $r$ raises the number of tokens that need to be updated on the GPU and reduces the number of KVs loaded from the external cache. These costs vary in opposite directions with $r$, and their relative magnitudes depend on the model, GPU, and cache path. Reuse has a low access cost in CPU memory; on SSDs/HDDs or other external storage media, increasing recomputation may reduce TTFT by replacing expensive reads. Section~\ref{sec:4.3} therefore adopts a scheduling method that combines analytical estimation with empirical calibration: it first locates the region where computation and transfer are balanced, then determines the deployment ratio from measured latency for complete requests.

% Source Word paragraph 203

\subsection{Computation and Transfer Cost Model}\label{sec:C.1}
% Source Word paragraph 204
Consider processing $N$ reusable cached tokens through $L$ Transformer layers. Given ratio $r$, each layer recomputes $rN$ tokens and loads the remaining $(1-r)N$ KV pairs. Let $t_c$ denote the effective per-token, per-layer recomputation cost, which is mainly affected by model width and GPU compute capability; $t_i$ denotes the effective loading cost of the corresponding KVs, jointly reflecting cache-medium bandwidth, DMA granularity, and PCIe utilization; and $t_o$ denotes fixed per-layer overhead, including kernel launches, CUDA event synchronization, and buffer management. These quantities are estimated during deployment initialization by profiling the target model and cache path.

% Source Word paragraph 205
The model treats $t_c$, $t_i$, and $t_o$ as amortized constants for a given configuration and uses a steady-state pipeline approximation that ignores startup and drain bubbles. The per-layer times of the computation and transfer paths are, respectively:
% Source Word paragraph 206
\begin{equation*}T_c(r)=rNt_c,\qquad T_i(r)=(1-r)Nt_i\tag{C.1}\label{eq:C.1}\end{equation*}
% Source Word paragraph 207
In the cross-layer pipeline in Appendix~\ref{sec:B}, cache transfer for the next layer overlaps with computation at the current layer, so the per-layer execution pace is determined by the slower of the two paths. Including the fixed overhead gives:
% Source Word paragraph 208
\begin{equation*}T_{\mathrm{layer}}(r)\approx\max\{rNt_c,(1-r)Nt_i\}+t_o\tag{C.2}\label{eq:C.2}\end{equation*}
% Source Word paragraph 209
Summing over $L$ layers yields the latency approximation used in Section~\ref{sec:4.3}:
% Source Word paragraph 210
\begin{equation*}T_{\mathrm{TTFT}}(r)\approx L\cdot\max\{rNt_c,(1-r)Nt_i\}+Lt_o\tag{C.3}\label{eq:C.3}\end{equation*}
% Source Word paragraph 211
The model defines two operating regions. When $r$ is small, cache transfer dominates, and increasing $r$ reduces the KVs waiting to be loaded; when $r$ is large, GPU recomputation dominates, and further increasing $r$ lengthens the computation path. The two paths intersect at $rNt_c=(1-r)Nt_i$, giving the analytical balance point:
% Source Word paragraph 212
\begin{equation*}r_0=\frac{t_i}{t_c+t_i}\tag{C.4}\label{eq:C.4}\end{equation*}
% Source Word paragraph 213
Thus, a higher cache access cost shifts the balance point toward a larger recomputation ratio, whereas a higher per-unit recomputation cost makes greater cache reuse more attractive. $N$ and $L$ determine the total workload, while $t_c$ and $t_i$ characterize the relative costs of the target configuration. The actual search uses TTFT for complete requests, including computation for new queries and uncached content, transfer startup, synchronization, and pipeline bubbles in the measurements.

% Source Word paragraph 214

\subsection{Warm-Started Golden-Section Search}\label{sec:C.2}
% Source Word paragraph 215
Calibration first establishes the allowed recomputation interval $[r_{\min},r_{\max}]$. Based on the quality--latency experiments in Appendix~\ref{sec:E.4}, we use $r_{\min}=15\%$ as an empirical lower bound for quality preservation and set $r_{\max}\leq 1$. This lower bound reserves a sufficient cross-chunk repair budget when optimizing latency on fast media. The analytical balance point is then clipped to this interval to serve as the initial probe for empirical search:
% Source Word paragraph 216
\begin{equation*}r_{\mathrm{init}}=\min\{r_{\max},\max\{r_{\min},r_0\}\}\tag{C.5}\label{eq:C.5}\end{equation*}
% Source Word paragraph 217
Let $\mathcal S_{\mathrm{cal}}$ be a calibration sample set with representative request lengths. Each candidate ratio selects recomputation positions along the same offline ranking and runs the complete prefill process on these requests. The calibration objective is to minimize average TTFT:
% Source Word paragraph 218
\begin{equation*}F(r)=\frac{1}{|\mathcal S_{\mathrm{cal}}|}\sum_{s\in\mathcal S_{\mathrm{cal}}}\operatorname{TTFT}(s;r)\tag{C.6}\label{eq:C.6}\end{equation*}
% Source Word paragraph 219
\begin{equation*}r^*=\mathop{\arg\min}_{r\in[r_{\min},r_{\max}]}F(r)\tag{C.7}\label{eq:C.7}\end{equation*}
% Source Word paragraph 220
The computation--transfer balance model has a unimodal structure that first decreases and then increases, providing a basis for one-dimensional derivative-free search. CacheTune uses golden-section search \citep{kochenderfer2019algorithms}, warm-started from the analytical initial value. Specifically, it first pairs $r_{\mathrm{init}}$ from Equation~\eqref{eq:C.5} with another probe in the interval to form an ordered pair, compares their measured average TTFT, and retains the more favorable subinterval; it then reinitializes the standard golden-section points within that subinterval. Algorithm~\ref{alg:1} denotes this clipped initial value by $r_0$.

% Source Word paragraph 221
In the standard update phase, let $\varphi=(\sqrt{5}-1)/2$ and choose $x_1=b-\varphi(b-a)$ and $x_2=a+\varphi(b-a)$ within the current interval $[a,b]$. After comparing $F(x_1)$ and $F(x_2)$, the search retains the interval on the lower-latency side and reuses one existing probe and its function value, requiring only one new candidate-ratio evaluation per iteration. When the interval width is less than $\varepsilon$, the algorithm returns the midpoint of the remaining interval as the deployment ratio. The analytical model supplies the starting point, while the empirical objective feeds the actual costs of transfer and pipelining into the choice of ratio.

\begin{algorithm}[tbp]
\caption{Hardware-Aware Calibration of the Recomputation Ratio}
\label{alg:1}
\small
\begin{tabularx}{\linewidth}{@{}r@{\quad}>{\raggedright\arraybackslash}X@{}}
& Input: calibration set $\mathcal S_{\mathrm{cal}}$; model $M$;
roofline prior $r_0$, obtained by clipping $t_i/(t_c+t_i)$
to $[r_{\min},r_{\max}]$; semantic constraint bounds
$r_{\min}$ and $r_{\max}$, satisfying $r_{\min}\leq r_{\max}$;
tolerance threshold $\varepsilon$. \\
& Output: selected recomputation ratio $r^*$. \\

1 & $F(x) \triangleq \operatorname{EvalTTFT}(M,\mathcal S_{\mathrm{cal}},x)$
\quad $\triangleright$\, Average TTFT on the calibration set \\

2 & $a \leftarrow r_{\min}$; $b \leftarrow r_{\max}$;
$\varphi \leftarrow (\sqrt{5}-1)/2$
\quad $\triangleright$\, Initialize the search \\

3 & If $r_0\leq(a+b)/2$, then
$(x_1,x_2) \leftarrow (r_0,a+\varphi(b-a))$; \\
& otherwise $(x_1,x_2) \leftarrow (b-\varphi(b-a),r_0)$
\quad $\triangleright$\, Construct ordered warm-start probes \\

4 & $(f_1,f_2) \leftarrow (F(x_1),F(x_2))$ \\

5 & If $f_1\leq f_2$, then $(a,b) \leftarrow (a,x_2)$;
otherwise $(a,b) \leftarrow (x_1,b)$ \\
& $\triangleright$\, Shrink the interval using unimodality \\

6 & $x_1 \leftarrow b-\varphi(b-a)$;
$x_2 \leftarrow a+\varphi(b-a)$
\quad $\triangleright$\, Reinitialize standard GSS \\

7 & $(f_1,f_2) \leftarrow (F(x_1),F(x_2))$ \\

8 & While $b-a\geq\varepsilon$: \\

9 & \hspace*{1em}If $f_1<f_2$: \\

10 & \hspace*{2em}$b \leftarrow x_2$;
$x_2 \leftarrow x_1$; $f_2 \leftarrow f_1$ \\

11 & \hspace*{2em}$x_1 \leftarrow b-\varphi(b-a)$;
$f_1 \leftarrow F(x_1)$ \\

12 & \hspace*{1em}Else: \\

13 & \hspace*{2em}$a \leftarrow x_1$;
$x_1 \leftarrow x_2$; $f_1 \leftarrow f_2$ \\

14 & \hspace*{2em}$x_2 \leftarrow a+\varphi(b-a)$;
$f_2 \leftarrow F(x_2)$ \\

15 & \hspace*{1em}End if \\

16 & End while \\

17 & Return $(a+b)/2$ \\
\end{tabularx}
\end{algorithm}
% Source Word paragraph 244
During the golden-section phase, the interval width shrinks by a factor of $\varphi$ in each iteration. If its initial width is $\Delta r$, reaching a ratio tolerance of $\varepsilon$ requires $O(\log(\Delta r/\varepsilon))$ iterations. Each candidate ratio is evaluated on $|\mathcal S_{\mathrm{cal}}|$ requests, so this phase requires $O(|\mathcal S_{\mathrm{cal}}|\cdot\log(\Delta r/\varepsilon))$ TTFT measurements, plus a fixed number of initial probes. The entire process is performed during deployment calibration, and the resulting ratio is used for subsequent requests. Online execution directly uses this ratio to select token importance indices and generate the complementary KV loading plan.

% Source Word paragraph 245

\section{Experimental Setup and Evaluation Methodology}\label{sec:D}
% Source Word paragraph 246
We evaluate CacheTune in terms of generation quality and prefill efficiency. The quality experiments examine how much cross-chunk information selective recomputation can recover after independently cached chunks are concatenated into a new context. The performance experiments examine the combined overhead of repair operations and cache transfer, comparing time to first token and serving throughput across storage tiers.

% Source Word paragraph 247

\subsection{Models, Hardware, and Software Environment}\label{sec:D.1}
% Source Word paragraph 248
We use three models, Mistral v0.3-7B, Llama 3-8B, and Qwen 2.5-32B \citep{DBLP:journals/corr/abs-2310-06825,DBLP:journals/corr/abs-2407-21783,qwen2024qwen25}, to evaluate CacheTune across model sizes. The hardware comprises two platforms: one with 2$\times$NVIDIA A100-40GB GPUs and another with 2$\times$NVIDIA GeForce RTX 4090-24GB GPUs. Both platforms connect to the host through PCIe Gen3$\times$16. Reusable KV Caches are placed in GPU memory, CPU memory, and disk cache pools. The disk experiments use an HDD cache pool on the A100 platform and an SSD cache pool on the RTX 4090 platform. These configurations provide different GPU computing capabilities and external-cache access costs for hardware-aware scheduling.
%%Sequential read/write measurements with fio give effective read and write bandwidths of approximately 205 MB/s and 201 MB/s for the HDD offloading path, respectively. The corresponding bandwidths for the SSD path are approximately 535 MB/s and 445 MB/s.

% Source Word paragraph 250
All experiments run on Ubuntu 22.04.1 with Linux kernel 6.8.0, PyTorch 2.2.1, and CUDA 12.4.

% Source Word paragraph 251

\subsection{Data Preparation}\label{sec:D.2}
% Source Word paragraph 252
The experiments cover five long-context datasets, comprising two summarization tasks and three multi-hop question-answering tasks. SAMSum constructs summarization contexts by concatenating multi-turn dialogues, while Multi-News constructs multi-document summarization contexts by concatenating news articles. MuSiQue, WikiMQA, and HotpotQA combine retrieved fragments from multiple documents into question-answering contexts. All these tasks require the model to use information from multiple reusable segments, and multi-hop question answering additionally requires connecting evidence distributed across segments. Table~\ref{tab:2} summarizes the task type, context construction, and evaluation metric for each dataset.

% Source Word paragraph 253
\begin{table}[htbp]
\centering
\caption{Summary of evaluation datasets and metrics.}
\label{tab:2}
\small
\begin{tabularx}{\linewidth}{@{}>{\raggedright\arraybackslash}p{.32\linewidth} >{\raggedright\arraybackslash}X l@{}}
\toprule
Dataset & Task and Context Construction & Metric $\uparrow$ \\
\midrule
SAMSum \citep{gliwa2019samsum} & Multi-turn dialogue summarization; concatenated dialogues & ROUGE-L \\
MuSiQue \citep{trivedi2022musique} & Multi-hop question answering; concatenated retrieved chunks & F1 \\
WikiMQA \citep{ho2020constructing} & Multi-hop question answering; concatenated retrieved chunks & F1 \\
HotpotQA \citep{DBLP:conf/emnlp/Yang0ZBCSM18} & Multi-hop question answering; concatenated retrieved chunks & F1 \\
Multi-News \citep{fabbri2019multi} & Multi-document summarization; concatenated news articles & ROUGE-L \\
\bottomrule
\end{tabularx}
\end{table}
% Source Word paragraph 272
SAMSum and Multi-News use ROUGE-L to evaluate summarization quality \citep{lin2004rouge}; MuSiQue, WikiMQA, and HotpotQA use answer F1. For both metrics, higher values indicate closer agreement between the generated output and the reference answer. Methods are compared on the same model and task, with Full Recompute providing a quality reference based on full-context encoding.

% Source Word paragraph 273

\subsection{Comparison Methods, Metrics, and Cache Placement}\label{sec:D.3}
% Source Word paragraph 274
We use eight baselines. Full Recompute performs standard prefill over the entire prompt; Full KV Reuse directly concatenates independently encoded chunk caches. vLLM Prefix Caching uses native Automatic Prefix Caching to reuse only strictly matching prefixes, recomputing the remaining content through the default prefill path. These three methods correspond to full-context computation, direct non-prefix reuse, and strict prefix reuse, respectively.

% Source Word paragraph 275
The remaining baselines restore cross-chunk dependencies in different ways. Following the released CacheBlend implementation, we use the first layer as the check layer to identify HKVD tokens from KV deviations, and then selectively recompute the same selected 15\% of positions in subsequent layers. EPIC updates the attention-sink positions at the beginning of each chunk, set to the first 16 tokens in our experiments. KVLink inserts 5 trained link tokens at each chunk boundary. For CacheSlide, we treat reusable documents as fixed segments and queries as updated segments. We use Weighted Correction Attention for selective recomputation and cache fusion, adapting it to the multi-chunk contexts in our experiments. We use an adapted implementation of ProphetKV based on its original method. In the first stage, query tokens are processed through successive layers to compute their attention to context tokens in the reused cache, and the scores are averaged over query positions and layers. In the second stage, the aggregated scores determine a shared token set whose KVs are recomputed at each layer, while the remaining KVs are reused directly. For a fair comparison, we also set ProphetKV's recomputation ratio to 15\%.

% Source Word paragraph 276
CacheTune uses a low-frequency cutoff ratio of $\alpha=0.5$ and a recomputation ratio of $r=15\%$ in the main CPU-cache experiments. Offline ranking determines which positions to update; online execution recomputes the KV pairs at these positions and loads the cache for the remaining positions. Appendix~\ref{sec:E.4} varies $\alpha$ and $r$ separately to analyze the effects of the spectral selection range and recomputation budget. The external-disk experiments additionally use calibrated recomputation ratios.

% Source Word paragraph 277
System performance is measured primarily by time to first token (TTFT) to assess the effect of prefill overhead on response latency. Figure~\ref{fig:3} jointly compares task quality and TTFT. Methods without multi-tier offloading optimizations, such as EPIC and KVLink, use GPU-resident caches, while CacheTune places reusable chunks in CPU memory and enables the sparse offloading pipeline. Figure~\ref{fig:4} further varies the request rate to examine how TTFT changes with increasing load and compare serving throughput across methods.

% Source Word paragraph 278
The storage-tier experiments compare methods using the same cache medium. In Table~\ref{tab:1}, CacheTune, CacheBlend, CacheSlide, and Prefix Caching are each run with GPU and CPU cache pools. CacheTune's GPU configuration uses the native cache reuse path, whereas its CPU configuration enables sparse transfer, asynchronous prefetching, and deferred RoPE restoration. Both use a fixed recomputation ratio of 15\%. Figure~\ref{fig:6} compares fixed and adaptive ratios on the A100/HDD and RTX 4090/SSD configurations, with TTFT speedups measured relative to Full Recompute on the corresponding configuration.

% Source Word paragraph 279

\subsection{Deployment Calibration of the Recomputation Ratio}\label{sec:D.4}
% % Source Word paragraph 280
% Hardware-aware calibration first profiles the target model and cache path to estimate computation and transfer costs, then initializes the ratio search at the analytical balance point. We use the first 10 SAMSum samples as the calibration set and select the recomputation ratio based on mean TTFT. The samples provide request lengths for latency measurements. Each candidate ratio runs the same calibration requests, allowing comparison of the actual computation, transfer, and synchronization overheads.

% % Source Word paragraph 281
% On the tested hardware and storage configurations, 5--10 samples yield relatively stable ratios, and profiling and search together take approximately 3--4 minutes. Calibration runs once when deploying to new hardware or a new cache medium. The selected ratio is then applied to reusable chunks and used for subsequent requests. The fast-cache experiments use the quality-preserving lower bound of $r=15\%$; the disk experiments select 36.4\% for the HDD configuration and 30.9\% for the SSD configuration. The search procedure is described in Appendix~\ref{sec:C}, and the corresponding latency results are presented in Appendix~\ref{sec:E.5}.
We use the first 10 SAMSum samples as the calibration set. Calibration measures computation, transfer, and synchronization costs without using reference answers or task-specific quality scores. Other datasets with comparable cache reuse patterns can therefore also serve as calibration workloads. Each candidate ratio is evaluated on the same calibration requests, and mean TTFT is used as the search objective. Hardware profiling and ratio search follow the procedure described in Appendix~\ref{sec:C}.

The fast-cache experiments use $r=15\%$, which also serves as the empirical lower bound for disk calibration. The disk experiments use the calibrated ratios of $r=36.4\%$ for A100/HDD and $r=30.9\%$ for RTX 4090/SSD. These ratios remain fixed during subsequent evaluation on each configuration. The corresponding latency results are reported in Appendix~\ref{sec:E.5}.

% Source Word paragraph 282

\section{Supplementary Experiments and Analysis}\label{sec:E}
% Source Word paragraph 283
CacheTune's frequency selection and hardware scheduling determine the positions and budget for recomputation, respectively. Our experiments examine these two decisions: spectral distributions and parameter studies assess the basis of the positional importance ranking; reuse of fixed indices tests recovery across contexts; and measurements across cache granularities and storage tiers examine how the recomputation budget affects data transfer and execution latency.

% Source Word paragraph 284

\subsection{Frequency-Domain Energy Distribution of the KV Cache}\label{sec:E.1}
\begin{figure}[tbp]
\centering
\includegraphics[width=1.0\linewidth]{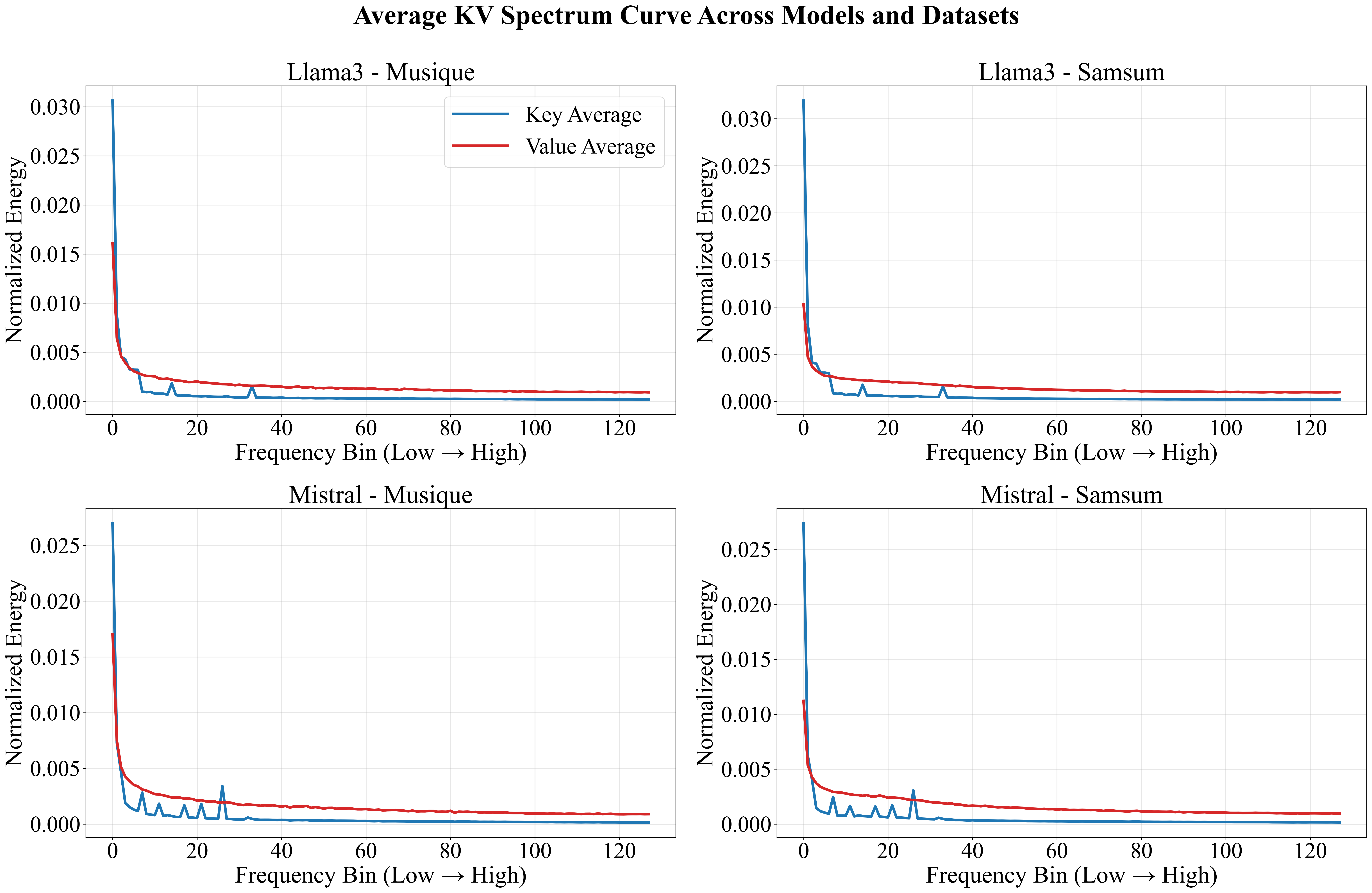}
\caption{Frequency-domain energy distribution of the KV Cache along the sequence dimension.}
\label{fig:9}
\end{figure}

% Source Word paragraph 285
Frequency-guided selection starts from the premise that KV sequences contain structure that can characterize token importance. To examine this structure, we run MuSiQue and SAMSum on Mistral v0.3-7B and Llama 3-8B, apply the fast Fourier transform (FFT) \citep{williamson2012discrete} separately to Keys and Values along the token dimension, and measure the energy fractions in different frequency bands. Low-frequency components correspond to slower variations along the sequence, whereas high-frequency components correspond to faster local variations. Figure~\ref{fig:9} presents the measurements for the two models and two task types.

% Source Word paragraph 286
% Source Word paragraph 287
% Source Word paragraph 288
All groups show a consistent concentration at low frequencies: the high-frequency energy fractions of Keys and Values decrease as frequency increases, and the lowest 20\% of the spectrum carries the vast majority of the total energy. This indicates that the main variations in KVs along the sequence can be compactly described by low-frequency structure. The phenomenon occurs in both summarization and multi-hop question answering, providing a basis for constructing an offline ranking from the frequency-domain features of each chunk itself.

% Source Word paragraph 289
CacheTune applies the inverse transform to the retained spectrum to recover a low-frequency signal aligned with the original sequence positions, then computes importance from the reconstructed magnitude at each token. The overall structure in the spectrum is thus converted into a ranking of positions that online recomputation can use directly. The spectral measurements motivate the use of low-frequency signals.

% Source Word paragraph 290

\subsection{Basis for Low-Frequency Selection and Robustness of Reuse across Contexts}\label{sec:E.2}
% Source Word paragraph 291
Independent encoding omits cross-chunk context, causing the KVs at some positions to deviate from full encoding. Using these positions with larger changes as a reference, we examine whether low-frequency reconstruction magnitudes can provide an effective KV importance ranking for recomputation, and further test how well the ranking recovers attention distributions and final answers.

% Source Word paragraph 292
Using Mistral v0.3-7B on SAMSum and MuSiQue, we compare the Keys and Values of the same token under independent chunk encoding and full-context encoding. Drift is computed from KV differences normalized within each layer, and Keys are compared before RoPE to exclude the effect of positional rotation. We use the 15\% of tokens with the largest drift as the reference set. The low-frequency, high-frequency, and random strategies each also select 15\% of positions; Recall@15\% is the intersection of the two sets divided by the size of the high-drift set.

% Source Word paragraph 293
\begin{table}[htbp]
\centering
\caption{Coverage of high-drift KVs by different selection strategies. Only 15\% of tokens are selected; Recall@15\% is reported.}
\label{tab:3}
\small
\begin{tabular}{lrr}
\toprule
Metric & SAMSum & MuSiQue \\
\midrule
LF Recall@15\% & 90.23\% & 86.31\% \\
HF Recall@15\% & 37.64\% & 24.90\% \\
Random Recall@15\% & 15.92\% & 14.62\% \\
\bottomrule
\end{tabular}
\end{table}
% Source Word paragraph 306
In Table~\ref{tab:3}, low-frequency selection achieves Recall@15\% of 90.23\% and 86.31\% on the two tasks, exceeding 37.64\% and 24.90\% for high-frequency selection and 15.92\% and 14.62\% for random selection. This indicates that low-frequency reconstruction magnitudes can identify, from independently cached representations, positions that change substantially in the full context. We further recompute the positions selected by the low-frequency strategy to test whether updating these KVs brings attention closer to the result of full-context encoding.

% Source Word paragraph 307
On SAMSum and MuSiQue, we use the attention distribution from full recomputation as the reference and compare its cosine similarity with those from full KV reuse, low-frequency-guided recomputation, and high-frequency-guided recomputation. Both the low-frequency and high-frequency strategies select 15\% of positions from each chunk, update their KVs in the composed context, and directly reuse the remaining positions. This allows attention recovery from different position selections to be compared under the same recomputation budget. Figure~\ref{fig:1}(a) presents the results across different queries and numbers of chunks.

% Source Word paragraph 308
Figure~\ref{fig:1}(a) shows that directly concatenating independently cached chunks leaves the attention distribution consistently different from that of full recomputation. The high-frequency strategy alleviates some of this deviation, whereas low-frequency recomputation maintains higher and more stable similarity across different queries and numbers of chunks. The representative heatmaps in Figure~\ref{fig:1}(b) show the same trend: full reuse and the high-frequency strategy exhibit clear shifts in attention positions, while the low-frequency strategy recovers a distribution closer to the reference. The preceding coverage experiment shows that the low-frequency ranking identifies high-drift positions; the attention results here further show that updating KVs according to this ranking helps restore the use of cross-chunk information.

% Source Word paragraph 309
To test whether this recomputation strategy improves final answers and whether offline indices remain effective in new contexts, we fix the cached document chunks and their within-chunk indices on MuSiQue, reuse them across different queries, and construct four context settings: the baseline order, random reordering, addition of a 1,024-token preceding context, and reordering followed by the addition of preceding context. When the global positions of chunks change, RoPE is applied to both reused and recomputed Keys using their corresponding new positions. The low-frequency, high-frequency, and random strategies all recompute 15\% of document tokens, with full KV reuse and full recomputation serving as references for direct reuse and full encoding, respectively. Figure~\ref{fig:1}(c) presents the overall results. The low-frequency strategy achieves the highest reuse quality in all four settings. In the order shown in Figure~\ref{fig:1}(c), its F1 differs from the corresponding full-recomputation result by 2.94, 1.98, 1.86, and 3.63 points, respectively, and improves over random selection by 8.21, 5.10, 8.14, and 5.54 points. In particular, after chunks are reordered or preceding context is added, retaining the existing indices preserves its advantage over the high-frequency, random, and full-reuse strategies. This indicates that positions identified from chunk content still help recover missing dependencies in new contexts.

% Source Word paragraph 310
Low-frequency reconstruction magnitudes provide a basis for selecting context-sensitive positions, and recomputing these positions brings attention and answer quality closer to full encoding. Frozen indices retain these benefits after changes to the query and layout, making this signal suitable for offline extraction during independent chunk encoding and subsequent use with the cache in later requests.

% Source Word paragraph 311

\subsection{Impact of Storage Tiers}\label{sec:E.3}
% Source Word paragraph 312
Cached chunks are both the content units of non-prefix reuse and the basic units of organization and access in external storage. As chunk size increases, full recomputation must process more tokens, while direct reuse must read more KVs. To examine how these two costs vary with chunk granularity, we measure the total overhead of recomputation and loading the complementary KVs from the cache pool under ideal conditions. Cache-loading costs are measured using a baseline implementation without optimizations such as cross-layer prefetching.

\begin{figure}[tbp]
\centering
\includegraphics[width=1.0\linewidth]{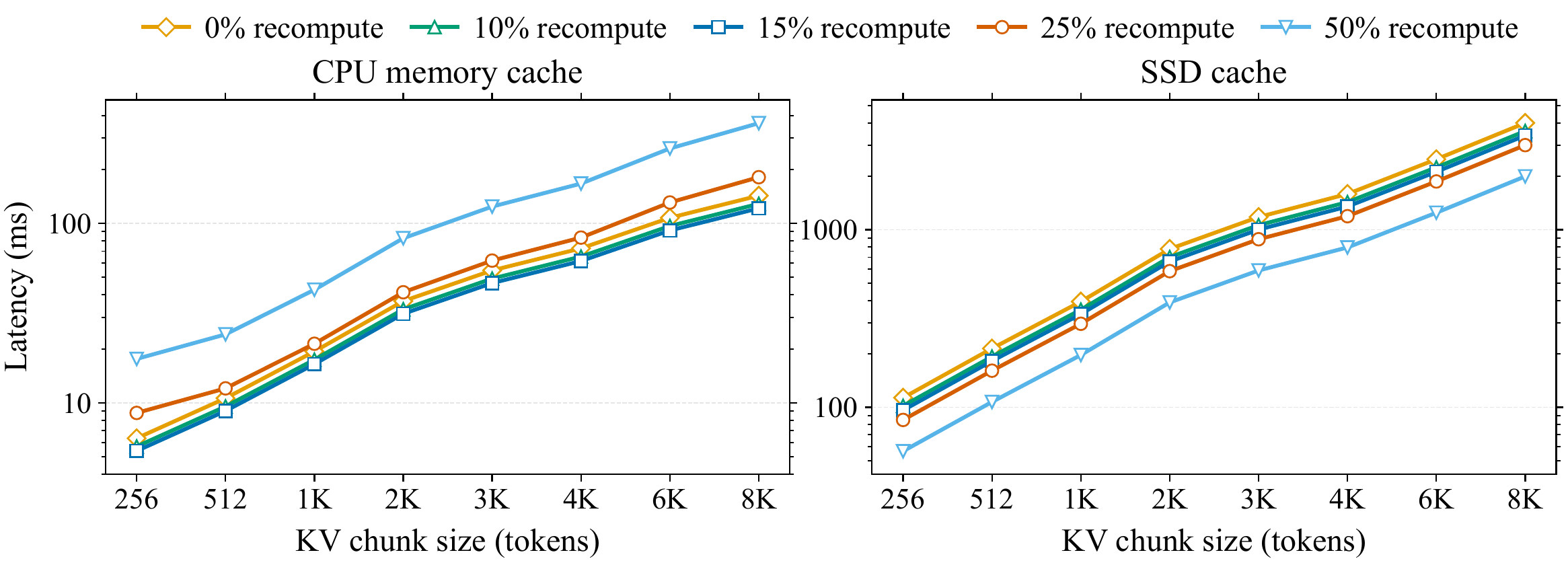}
\caption{Effect of the recomputation ratio on KV Cache reuse latency.}
\label{fig:10}
\end{figure}

Figure~\ref{fig:10} shows that the absolute latency of each path increases with chunk length, while the relative performance of different recomputation ratios depends on the cache medium. CPU memory has a low access cost, so a small recomputation ratio keeps latency low: most KVs come from the cache, and only selected positions incur online computation. Further increasing the recomputation ratio adds GPU work and gradually reduces the computational savings from reuse.

Reading data from the SSD is more expensive, shifting the balance between computation and transfer. Increasing the recomputation ratio reduces the complementary KV portion that must be read and moves more work to the GPU. The savings in read overhead can offset the additional computation cost. In this configuration, higher recomputation ratios yield lower cache reuse latency than full reuse.

The two sets of curves show that the benefits of cache reuse depend jointly on chunk length, GPU computation cost, and cache access cost. Low recomputation ratios suit fast memory, whereas slower storage benefits more from replacing data movement with computation. Accordingly, the scheduler in Section~\ref{sec:4.3} profiles the target storage path and selects the recomputation ratio based on measured TTFT, applying the trade-off observed at chunk granularity to end-to-end request execution.

\subsection{Effects of the Frequency Cutoff Ratio and Recomputation Ratio}\label{sec:E.4}
\begin{figure}[tbp]
\centering
\includegraphics[width=1.0\linewidth]{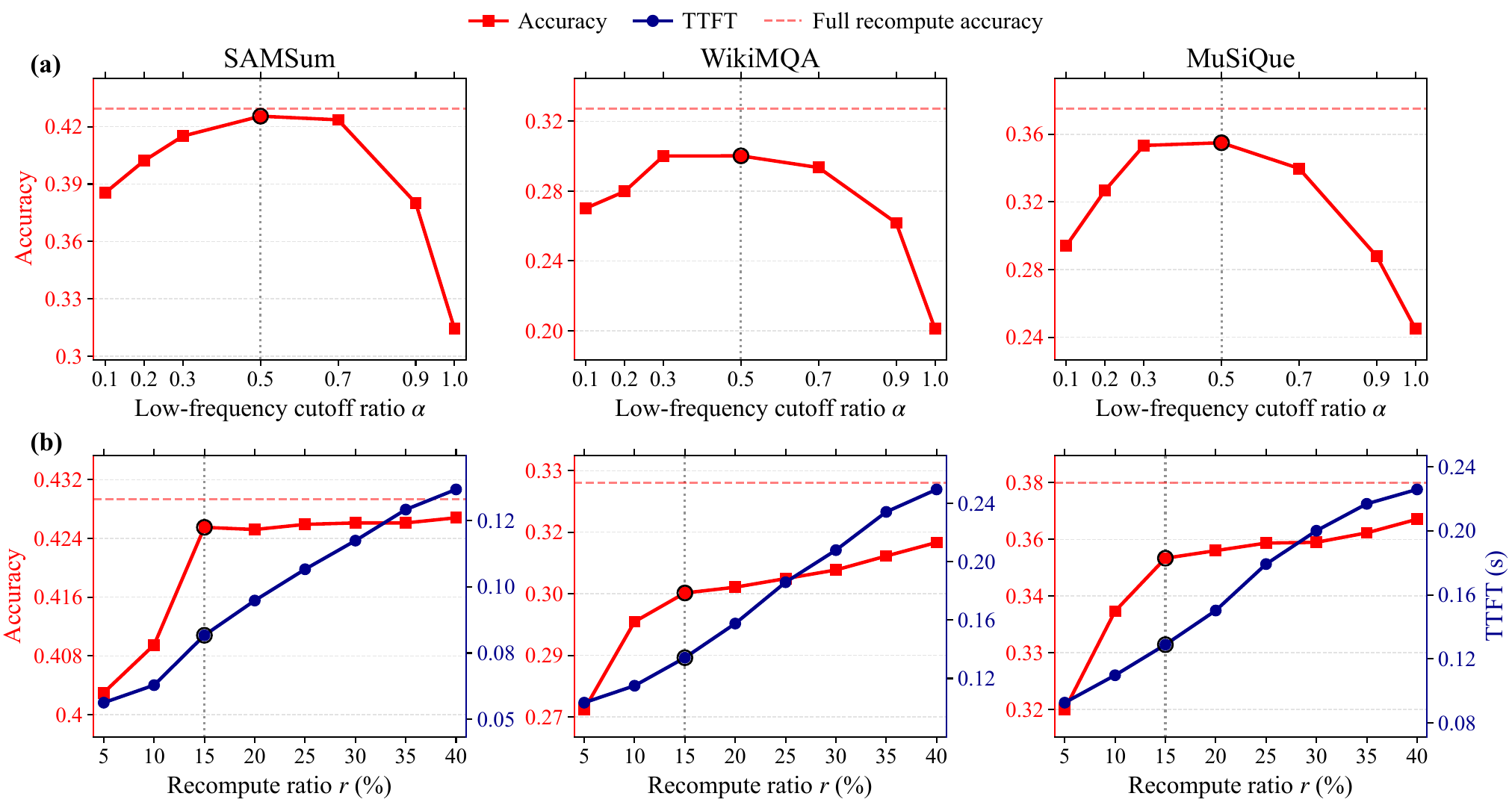}
\caption{Sensitivity of model accuracy and TTFT to the low-frequency cutoff ratio $\alpha$ and recomputation ratio $r$, using Mistral v0.3-7B and a CPU memory KV Cache pool.}
\label{fig:11}
\end{figure}

% Source Word paragraph 319
The low-frequency cutoff ratio $\alpha$ determines which spectral components contribute to token scoring, while the recomputation ratio $r$ determines how many positions are updated along that ranking. Using Mistral v0.3-7B and a CPU memory KV Cache pool, we vary these two parameters separately on SAMSum, WikiMQA, and MuSiQue. Figure~\ref{fig:11}(a) fixes $r=15\%$ and measures task quality for different $\alpha$ values; Figure~\ref{fig:11}(b) fixes $\alpha=0.5$ and measures both quality and TTFT. The quality of full recomputation serves as the reference line.

% Source Word paragraph 320
% Source Word paragraph 321
% Source Word paragraph 322
We first increase $\alpha$ from 0.1 to 1.0. Generation quality improves substantially on all three datasets as $\alpha$ increases to 0.3--0.5. CacheTune maintains strong quality within $\alpha\in[0.3,0.7]$, with $\alpha=0.5$ providing the most stable performance across tasks. Quality declines when $\alpha$ is further increased to 0.9 or 1.0. A small cutoff ratio may filter out spectral information useful for ranking positions, while retaining too many high-frequency components weakens the role of low-frequency structure in scoring. The intermediate range therefore provides better recovery.

% Source Word paragraph 323
The setting $\alpha=1.0$ provides a useful comparison. All frequencies are retained, the inverse transform recovers the original KVs, and the importance scores approximately reduce to a ranking by the L2 norm of the original KV pairs. This setting underperforms $\alpha=0.5$ on all three tasks, indicating that filtering the spectrum improves the selection of recomputation positions. Given its stable performance across tasks, we use $\alpha=0.5$ as the default cutoff ratio.

% Source Word paragraph 324
We then fix $\alpha=0.5$ and increase $r$ from 5\% to 40\%. As more positions are updated in the current context, model quality generally improves and then gradually saturates, while TTFT continues to rise with the amount of online computation. A ratio of $r=15\%$ recovers most of the quality loss caused by non-prefix reuse while maintaining a TTFT speedup of more than $4\times$ over full recomputation. This ratio therefore provides a favorable quality--latency trade-off in the CPU cache setting and is used as the default in the main experiments.

% Source Word paragraph 325
This result also determines the quality lower bound for the hardware search. On fast media such as GPU and CPU memory, the low cache access cost places the latency balance point from the cost model in Section~\ref{sec:4.3} below 10\%. The quality curves show that an excessively small recomputation budget leaves more pronounced quality loss, so we set $r_{\min}$ to 15\%. For SSDs/HDDs, reading complementary KVs is more expensive, and increasing $r$ can reduce external transfer. Appendix~\ref{sec:E.5} further reports the calibrated ratios and end-to-end latency in these settings.

\subsection{Balancing Recomputation and Transfer on External Storage}
\label{sec:E.5}

External caching expands the storage space for reusable KVs but also introduces additional data access overhead. Using Mistral v0.3-7B, this section first compares end-to-end latency with GPU-resident and CPU-resident caches, then evaluates the three cache-loading optimizations introduced in Section~\ref{sec:4.2} through cumulative ablation. We subsequently extend the cache pool to SSDs/HDDs to examine whether adjusting the recomputation ratio can alleviate slower cache reads.

\textbf{GPU- and CPU-resident caches.} In the GPU--CPU comparison, CacheTune fixes $r=15\%$, and CacheTune, CacheBlend, CacheSlide, and prefix caching use the same medium within each group of experiments. CacheTune's GPU path uses the native cache reuse implementation, while its CPU path integrates sparse transfer, asynchronous prefetching, and deferred positional recovery into the same pipeline. Table~\ref{tab:1} shows that the optimized CPU path retains the low-latency benefits on all three tasks; compared with CacheBlend, CacheSlide, and prefix caching under the same CPU cache setting, it reduces average TTFT by approximately 25.3\%, 11.9\%, and 74.4\%, respectively. This result reflects the role of applying offline indices to cache loading. Knowing in advance which positions require recomputation, the system can skip reading these KVs and prefetch the complementary cache for subsequent layers while the current layer executes. By coordinating transfer, recomputation, and positional recovery through the pipeline, the system overlaps external cache access with computation wherever possible, allowing the capacity expansion provided by CPU caching to coexist with low TTFT.

\textbf{Cache-loading ablation.} To further distinguish the benefits of these optimizations, we fix $\alpha=0.5$ and $r=15\%$ under the same CPU cache configuration, using the same requests, cached chunks, and recomputation indices. Starting from a configuration without the cache-loading optimizations introduced in Section~\ref{sec:4.2}, we successively add pipeline overlap and cross-layer prefetching, sparse loading, and deferred RoPE. The final row corresponds to the full configuration with all three optimizations enabled.

\begin{table}[tbp]
    \centering
    \caption{TTFT (s) as cache-loading optimizations are added progressively. Each row adds the indicated optimization to the preceding configuration; the final row is the full configuration.}
    \label{tab:loading_ablation}
    \small
    \setlength{\tabcolsep}{7pt}
    \begin{tabular}{@{}lrrr@{}}
        \toprule
        Configuration
        & SAMSum
        & WikiMQA
        & MuSiQue \\
        \midrule
        No cache-loading optimizations
        & 0.179 & 0.289 & 0.281 \\
        + Pipeline overlap and cross-layer prefetching
        & 0.094 & 0.159 & 0.151 \\
        + Sparse loading
        & 0.086 & 0.140 & 0.137 \\
        + Deferred RoPE (full configuration)
        & 0.082 & 0.134 & 0.129 \\
        \bottomrule
    \end{tabular}
\end{table}

Pipeline overlap and cross-layer prefetching provide the largest latency reduction. Adding this optimization reduces TTFT from 179/289/281~ms to 94/159/151~ms across the three datasets, a reduction of 45.0\%--47.5\%. Computation in the current layer proceeds alongside cache loading for subsequent layers, placing previously serialized loading work within the computation window and reducing forward-pass stalls while waiting for KVs.

Adding sparse loading further reduces TTFT to 86/140/137~ms, an additional reduction of 8.5\%--11.9\% relative to the preceding configuration. The offline indices allow the system to skip KVs that will be overwritten by recomputation and transfer only the reuse complement. At $r=15\%$, the effective KV transfer volume is theoretically reduced by approximately 15\%. Because the pipeline already hides part of the transfer, this reduction in data volume does not translate proportionally into lower TTFT.

Finally, adding deferred RoPE reduces TTFT to 82/134/129~ms, a further reduction of 4.3\%--5.8\% relative to the preceding configuration. Global positional recovery for reused Keys is incorporated into the asynchronous execution path, reducing synchronization waits for positional correction in the main forward stream. This optimization changes when positional recovery is performed and provides a smaller benefit than the preceding two optimizations. With all three enabled, the full configuration reduces TTFT by 53.6\%--54.2\% compared with the configuration without cache-loading optimizations.

% Source Word paragraph 330
% \textbf{Calibration for SSD/HDD caches.} The disk experiments use two complete configurations: an A100 platform connected to an HDD cache pool, with sequential read/write bandwidth of approximately 205/201 MB/s, and an RTX 4090 platform connected to an SSD cache pool, with sequential read/write bandwidth of approximately 535/445 MB/s. On each configuration, we compare a fixed ratio of $r=15\%$ with hardware-aware calibration. Figure~\ref{fig:6} presents the TTFT of each method, and Table~\ref{tab:4} summarizes the average speedups with fixed and calibrated ratios.
\textbf{Calibration for SSD/HDD caches.} We compare a fixed ratio of $r=15\%$ with hardware-aware calibration. Figure~\ref{fig:6} presents the TTFT of each method, and Table~\ref{tab:4} summarizes the average speedups with fixed and calibrated ratios. With a fixed ratio of $r=15\%$, the HDD and SSD configurations achieve TTFT speedups of $1.92\times$ and $2.05\times$, respectively, over full recomputation in their corresponding configurations. Hardware-aware search selects $r=36.4\%$ for the HDD configuration and $r=30.9\%$ for the SSD configuration, increasing the speedups to $2.36\times$ and $2.34\times$, respectively. Both configurations reduce cache reads by increasing recomputation, showing that assigning more work to the GPU can improve end-to-end latency when disk access is expensive.
%The disk experiments use two complete configurations: an A100 platform connected to an HDD cache pool and an RTX 4090 platform connected to an SSD cache pool. On each configuration, 

%%Calibration uses the first 10 samples from SAMSum to adjust the ratio according to measured average TTFT, then applies the selected ratio to subsequent requests.

% Source Word paragraph 331
% \begin{table}[htbp]
% \centering
% \caption{Comparison of fixed and calibrated recomputation ratios. Speedups are relative to full recomputation in each configuration.}
% \label{tab:4}
% \small
% \begin{tabularx}{\linewidth}{@{}>{\raggedright\arraybackslash}p{.30\linewidth} *{4}{>{\centering\arraybackslash}X}@{}}
% \toprule
% Cache pool & Fixed $r$ & Speedup with fixed ratio & Calibrated $r$ & Speedup with calibrated ratio \\
% \midrule
% HDD & 15\% & $1.92\times$ & 36.4\% & $2.36\times$ \\
% SSD & 15\% & $2.05\times$ & 30.9\% & $2.34\times$ \\
% \bottomrule
% \end{tabularx}
% \end{table}

\begin{table}[htbp]
\centering
\caption{Comparison of fixed and calibrated recomputation ratios.
Speedups are relative to full recomputation in each configuration.}
\label{tab:4}
\small
\setlength{\tabcolsep}{8pt}
\renewcommand{\arraystretch}{1.05}
\begin{tabular}{@{}lcccc@{}}
\toprule
& \multicolumn{2}{c}{Fixed}
& \multicolumn{2}{c}{Calibrated} \\
\cmidrule(lr){2-3}\cmidrule(lr){4-5}
Cache pool & $r$ & Speedup & $r$ & Speedup \\
\midrule
HDD & 15\% & $1.92\times$ & 36.4\% & $2.36\times$ \\
SSD & 15\% & $2.05\times$ & 30.9\% & $2.34\times$ \\
\bottomrule
\end{tabular}
\end{table}
% Source Word paragraph 347

% Source Word paragraph 348
The selected ratios reflect the combined costs of computation and transfer in each configuration. A100/HDD and RTX 4090/SSD differ in GPU compute capability and storage bandwidth, and calibration seeks the balance between the two execution paths separately for each. The HDD configuration uses a higher ratio, moving more work away from the slower read path, whereas the SSD configuration retains more cache reuse. Compared with the 15\% used in the CPU memory setting, these two results further show that the recomputation budget serves both quality recovery and the reduction of slow I/O. The calibration of $r$ minimizes measured TTFT while keeping the empirical quality-preserving lower bound $r_{\min}$ fixed, without using reference answers. With the same model and hardware configuration, the calibrated ratio can be reused across tasks with similar workload characteristics without task-specific recalibration.

% Source Word paragraph 349
%%\Needspace{24\baselineskip}

\section{Discussion and Limitations}\label{sec:F}
Our experimental results show that CacheTune’s advantage stems from the tight coordination of its three key components. Unlike prior approaches that rely on heuristic rules, attention relationships, or learned signals to determine recomputation targets online, CacheTune constructs a query-independent token-importance ranking for each KV Chunk during cache generation based on its intrinsic frequency-domain characteristics. This ranking is used not only to identify the critical KVs that should be recomputed under the global context, but also directly as a cache-loading plan, allowing the system to transfer only the complementary KVs that will be reused. In this way, CacheTune transforms the conventional full-cache loading followed by partial repair workflow into a complementary KV-construction process based on selective recomputation and sparse transfer, thereby reducing I/O overhead. Under this execution model, the recomputation ratio jointly determines GPU computation and data-transfer overhead across cache media. The hardware-aware scheduler therefore adapts the workload allocation between the recomputation and transfer paths according to the compute and I/O characteristics of the target storage tier.
%%and the volume of KV Cache data stored in the external cache pool
%%external-cache footprint,

CacheTune’s cache-pool abstraction further has the potential to extend to emerging heterogeneous memory architectures. For example, with memory-interconnect technologies such as Compute Express Link (CXL) \citep{10.1145/3669900,seo2025oasis}, extended memory beyond the GPU and CPU can be organized as a more unified remote cache pool; CacheTune can still estimate the effective transfer cost through hardware profiling and adaptively tune the recomputation--reuse ratio. CacheTune nevertheless has several limitations. When external-storage bandwidth is extremely low or the asynchronous overlap window is insufficient, the system may remain I/O-bound. In future work, we will further evaluate CacheTune under higher concurrency, CXL-based disaggregated memory, and more complex workloads.

% Source Word paragraph 352

\end{document}